\documentclass[sigconf,nonacm]{acmart}
\usepackage{booktabs}
\usepackage{graphicx}
\usepackage{xcolor}
\usepackage{subcaption}
\usepackage{tikz}
\usepackage{tcolorbox}
\usepackage{multirow}
\usepackage{url}
\usepackage{makecell}
\usepackage{enumitem}
\usepackage{pifont}
\usepackage{bm}
\usepackage{float}
\usepackage{adjustbox}
\usepackage{colortbl}

\usepackage{xspace}
\usepackage{caption}

\newsavebox{\lefttablebox}
\newsavebox{\righttablebox}

\newcommand{\IL}{\texttt{CITADEL}\xspace}

\author{Aymen Bouferroum\textsuperscript{a},
Ildi Alla\textsuperscript{b},
Valeria Loscri\textsuperscript{a},
Abderrahim Benslimane\textsuperscript{c},
Vincent Lenders\textsuperscript{b}}
\affiliation{
  \institution{\textsuperscript{a}Inria Lille-Nord Europe, Lille, France}
  \institution{\textsuperscript{b}University of Luxembourg, Luxembourg}
  \institution{\textsuperscript{c}LIA/CERI, University of Avignon, Avignon, France}
  \country{}}

\renewcommand{\shortauthors}{Aymen Bouferroum et al.}

\title{\IL: CSI-Based Jamming Detection and Open-Set Classification for IIoT Networks}

\begin{abstract}
Radio frequency jamming poses a critical threat to the availability of wireless Industrial Internet of Things (IIoT) networks. Existing detection and classification techniques are poorly suited to this setting: coarse signal-strength and cross-layer features lack information richness, while raw I/Q baseband approaches require hardware and throughput that is impractical at the scale of hundred-node IIoT deployments.
This paper presents \IL, a lightweight two-stage hierarchical pipeline that uses only Channel State Information (CSI) measurements, which are natively available on commodity IIoT devices, to detect and classify jamming attacks including previously unseen ones. While prior work has shown that jamming leaves observable CSI signatures, \IL is the first system to translate this insight into an end-to-end pipeline that jointly achieves closed-set classification of known attacks, open-set detection of zero-day attacks, and resistance to adversarial evasion.
Evaluated across 6 known attack types and 15 zero-day scenarios, \IL achieves 100\% known-attack detection and 97.1\% zero-day detection at a 0.4\% end-to-end false positive rate. Under adversarial evaluation spanning white-box and black-box threat models, gradient-based evasion remains below 2\% across all tested perturbation budgets and the strongest published CSI attack generator achieves less than 5\% average evasion. A systematic comparison against eight baselines confirms that no existing method achieves comparable performance on CSI data across all three axes: detection, generalization, and robustness. The full pipeline completes inference in 14.2\,ms at 95.9\,mJ on an edge GPU, establishing \IL as a practical solution for large-scale IIoT network security.

\end{abstract}


\begin{document}
\maketitle

\section{Introduction}
\label{sec:introduction}

Jamming has become a major threat to Industrial Internet of Things (IIoT) ecosystems that wirelessly interconnect sensors, mobile human-machine interfaces, and condition-monitoring nodes alongside industrial control protocols~\cite{sadeghi2015security, serror2021challenges, kayan2022cybersecurity}. The broadcast nature of the wireless medium means that any adversary within radio range can inject interference to disrupt legitimate transmissions~\cite{xu2005feasibility, pirayesh2022jamming}. 
The widespread availability of low-cost commodity
software-defined radios~(SDRs), which can generate arbitrary waveforms with precise
timing and frequency control~\cite{ali2022jamrf}, has dramatically lowered the barrier
to mount such attacks, expanding both their sophistication and their variety.

The consequences can be severe. In semiconductor fabrication, a jammed condition-monitoring sensor that fails to report abnormal vibration may delay a safety interlock, allowing equipment damage to propagate~\cite{lee2014german}. In chemical processing, masked telemetry can conceal a runaway exotherm until manual intervention is too late~\cite{iaiani2021outage}. In autonomous logistics, disrupted control signals can halt conveyor systems or misroute automated guided vehicles~\cite{quarta2017experimental, serror2021challenges}. These are not \textit{hypothetical risks}. They reflect the operational dependencies that make IIoT
infrastructure an increasingly attractive target for adversaries.

\begin{figure}[t]
  \vspace{10pt}
  \centering
  \includegraphics[width=\columnwidth]{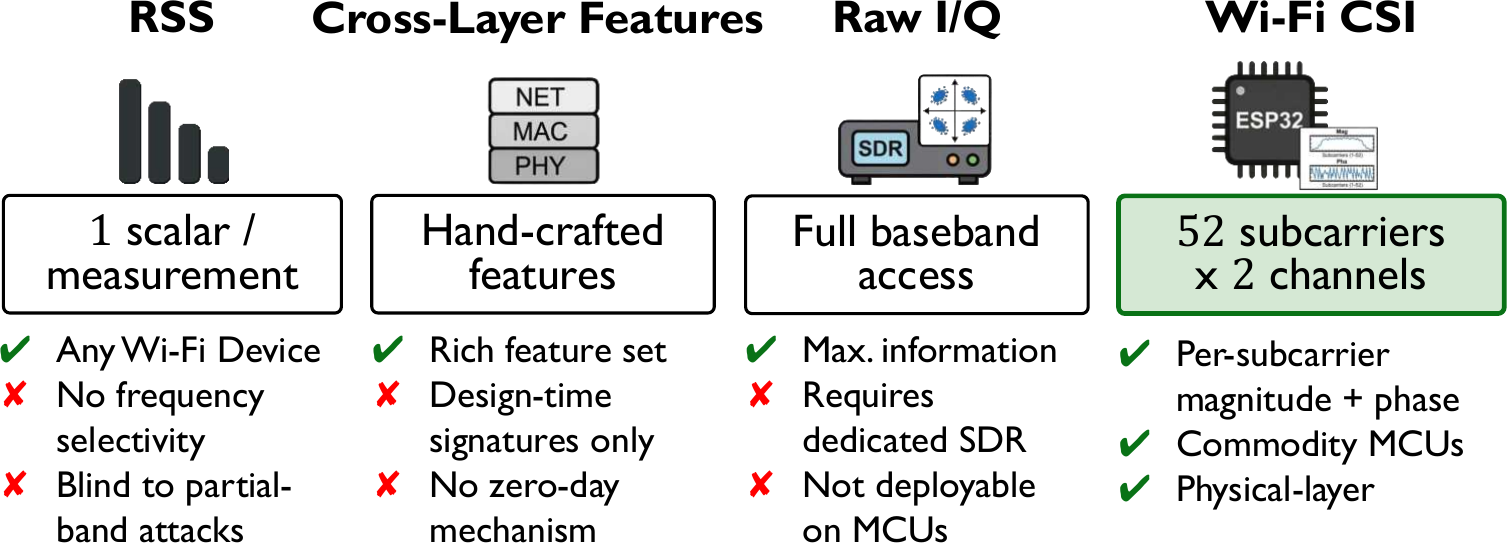}
  \caption{Comparison of four physical-layer sensing modalities for jamming detection. CSI uniquely combines per-subcarrier spectral resolution with commodity microcontroller availability, offering the strongest trade-off between information and real-time deployment.}
  \label{fig:modalities}
\end{figure}

Jamming detection and classification systems exist to counter these threats, yet
existing approaches are poorly suited to the IIoT setting. Techniques based on Received
Signal Strength~(RSS) reduce the entire wireless channel state to a single scalar per
measurement, discarding all frequency-selective structure. Although RSS is universally
available, it cannot distinguish jamming types, rendering it ineffective against
partial-band and sweeping interference~\cite{hussain2022}. Cross-layer approaches
aggregate protocol statistics into hand-crafted feature vectors, but these capture the
\textit{symptoms} of jamming rather than its physical signature, making them ill-suited
to detect attacks specifically designed to evade detection~\cite{punal2014wowmom, hachimi2020multistage,panitsas2025}. Raw I/Q
samples offer full baseband access and preserve maximal channel information, but they
require dedicated SDR receivers that are incompatible with commodity microcontrollers
and impractical for large-scale deployments where hundreds of low-cost sensor nodes must
be monitored simultaneously~\cite{sciancalepore2023}. Today, no existing technique jointly
satisfies the information richness, hardware compatibility, and scalability requirements
of real-world IIoT environments.

Among physical-layer sensing modalities, \textit{Channel State Information} (CSI) stands out as a compelling foundation for jamming detection in IIoT networks (see Figure~\ref{fig:modalities}). 
CSI captures per-subcarrier amplitude and phase across the OFDM frequency band, providing a spectrotemporal fingerprint far richer than aggregate metrics such as RSS, while remaining extractable at negligible cost from commodity Wi-Fi hardware, including microcontrollers, single-board computers, and off-the-shelf routers~\cite{halperin2011tool, Espressif_CSI_2025, gringoli2019free}. The physical invariant underlying CSI-based detection is that any RF-domain attack inevitably perturbs these measurements. An attacker cannot jam a link without adding energy that manifests in subcarrier magnitudes, phases, or both~\cite{ali2022jamrf, 11096274}.
Yet no prior work has translated this observation into a system that simultaneously achieves closed-set classification of known attacks, open-set detection of zero-day attacks, and resilience to adversarial evasion within the computational constraints of IIoT
deployments. These requirements motivate four research questions that guide the design and evaluation of this work:

\begin{tcolorbox}[colback=gray!8, colframe=gray!40, boxrule=0.4pt, arc=2pt, left=4pt, right=4pt, top=3pt, bottom=3pt]
\textbf{RQ1} (\textit{Input sufficiency}): Does CSI carry sufficient discriminative information to detect physical-layer jamming, or does reliable detection require richer input modalities?

\textbf{RQ2} (\textit{Zero-day generalization}): Can a CSI-based detector identify previously unseen jamming strategies without any zero-day data during model training or threshold calibration?

\textbf{RQ3} (\textit{Adversarial resilience}): Can such a detector withstand adversarial evasion attacks under physically realizable perturbation constraints?

\textbf{RQ4} (\textit{Edge deployment}): Can the detection pipeline operate in real time on commodity microcontrollers and edge GPUs within the computational hierarchy of IIoT environments?
\end{tcolorbox}

\begin{figure}[t]
  \centering
  \includegraphics[width=\columnwidth]{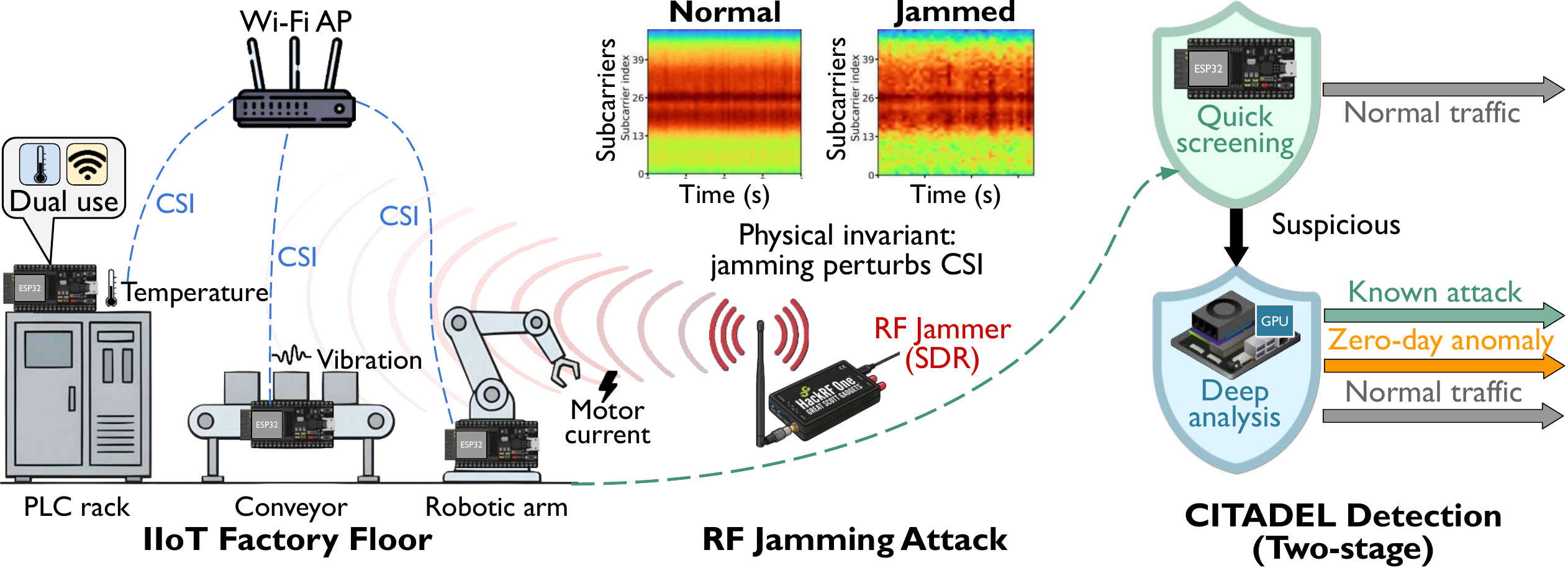}
  \caption{Deployment scenario. IIoT nodes extract per-subcarrier CSI from Wi-Fi frames, and an SDR jammer perturbs these measurements. \textsc{Citadel} detects attacks via a lightweight binary trigger on the nodes and a multi-signal ensemble on an edge GPU.}
  \label{fig:scenario}
\end{figure}

We present \textsc{Citadel}, a two-stage hierarchical detection system that addresses all four research questions through decomposition across hardware tiers. As illustrated in Figure~\ref{fig:scenario}, \textsc{Citadel} operates entirely on per-subcarrier CSI without requiring raw I/Q samples, dedicated spectrum analyzers, or infrastructure-level telemetry (\textbf{RQ1}). Stage~1 deploys a lightweight binary trigger on sensor nodes, screening benign traffic at sub-millisecond latency so that only suspicious windows reach the analysis backend (\textbf{RQ4}). Stage~2 runs on an edge GPU and fuses complementary anomaly signals operating in orthogonal information spaces to detect both known and previously unseen attacks without any zero-day training data (\textbf{RQ2}).
On six known jamming types at three power levels and 15 zero-day scenarios, \textsc{Citadel} achieves 100\% known-attack detection and 97.1\% zero-day anomaly detection at 0.4\% end-to-end (E2E) false positive rate (FPR). Under adversarial evaluation spanning white-box and black-box threat models, gradient-based evasion stays below 2\% at every tested perturbation budget, while Magmaw~\cite{magmaw2025}, the strongest published CSI attack generator achieves less than 5\% average evasion (\textbf{RQ3}).

\smallskip

\noindent In summary, our main contributions are:
\begin{itemize}[leftmargin=1.2em,itemsep=1pt,topsep=2pt]

\item We present \textsc{Citadel}, the \textit{first end-to-end CSI-based jamming detection system} for IIoT that jointly addresses known-attack classification, zero-day generalization, and adversarial resilience on commodity hardware. A two-stage hierarchical architecture decomposes detection across hardware tiers: a 1,362-parameter binary trigger on microcontrollers and a multi-signal out-of-distribution (OOD) ensemble on an edge GPU, completing end-to-end inference in 14.2\,ms at 95.9\,mJ.

\item We propose an \textit{OOD ensemble} that fuses three complementary anomaly signals (diffusion-based reconstruction divergence, classifier energy, and feature-space distance) calibrated exclusively on in-distribution data via K-fold cross-validation. In 18 known-attack scenarios and 15 zero-day scenarios, \textsc{Citadel} achieves 100\% known-attacks detection and 97.1\% zero-day anomaly detection respectively, and 0.4\% E2E FPR \textit{without} any zero-day data during training or threshold selection.

\item We conduct the \textit{first adversarial robustness} evaluation of a CSI-based jamming detector, 
spanning white-box gradient attacks, black-box transfer and query attacks, and three state-of-the-art perturbation generators. A comparative study against eight baseline methods confirms that no existing method matches \textsc{Citadel}'s joint performance.
 
\end{itemize}

\section{CSI Background and Detection Challenges}
\label{sec:background}

The CSI captures per-subcarrier amplitude and phase information from every transmission frame. In OFDM-based Wi-Fi, the receiver estimates the CSI as the complex channel response for each subcarrier $k$ at time~$t$:
\begin{equation}
H(k, t) = |H(k, t)| \cdot e^{j \angle H(k, t)},
\label{eq:csi}
\end{equation}
where $|H(k, t)|$ is the magnitude and $\angle H(k, t)$ is the phase shift imposed by the wireless channel~\cite{halperin2011tool}. Magnitude is sensitive to power-domain interference, while phase captures path-length perturbations and the discontinuities that broadband jamming introduces across subcarrier boundaries~\cite{zhang2023csi}. Unlike RSS, CSI preserves frequency selectivity across all subcarriers. Unlike I/Q, CSI is available on commodity microcontrollers at negligible cost. 
CSI thus offers a \textit{trade-off between information richness and deployment feasibility} (\textbf{RQ1}).

\begin{figure}[t]
  \centering
  \includegraphics[width=\columnwidth]{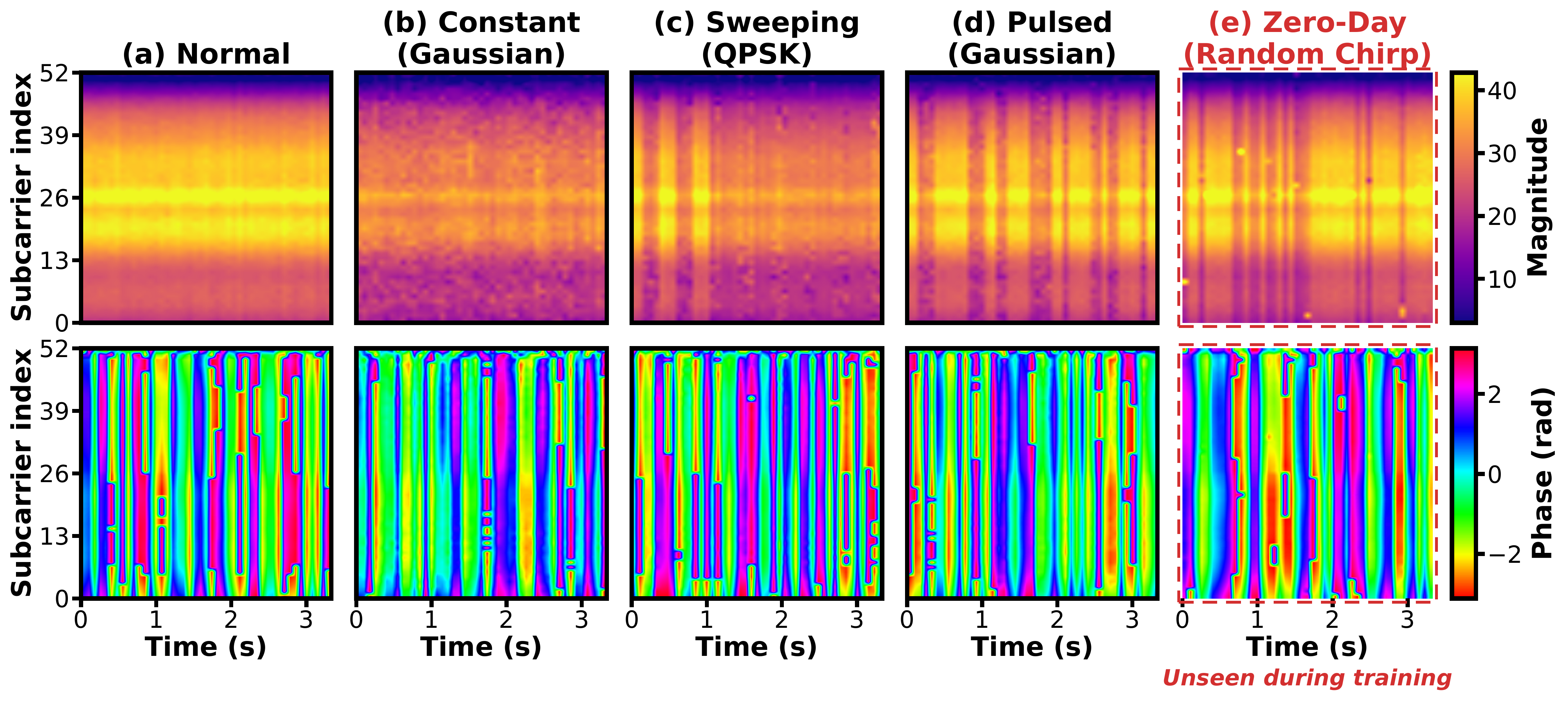}
\caption{CSI magnitude (top) and phase (bottom) spectrograms from our testbed. (a)~Normal traffic. (b)--(d)~Three known jamming strategies, each producing a visually distinct spectrotemporal signature. (e)~A zero-day scenario whose pattern partially overlaps with known classes, illustrating the generalization challenge.}
  \label{fig:csi_signatures}
\end{figure}

When a jammer emits RF interference, the jammer's signal $J(k,t)$ superposes onto the legitimate channel: $\tilde{H}(k,t) = H(k,t) + J(k,t)$~\cite{ali2022jamrf}.
Different jamming strategies produce distinct spectrotemporal patterns in CSI: constant jamming elevates magnitude uniformly across all subcarriers; sweeping jamming produces intermittent broadband bursts each time the sweep crosses the monitored channel's bandwidth, resembling pulsed interference but with timing governed by the sweep rate rather than a fixed duty cycle; and pulsed jamming introduces periodic high-energy bursts with quiet intervals between them~\cite{11096274}. 

Figure~\ref{fig:csi_signatures} visualizes both magnitude and phase signatures alongside normal traffic and a zero-day scenario from our testbed (Section~\ref{sec:implementation}). The waveform axis further modulates signatures: Gaussian noise produces broadband elevation, QPSK introduces structured modulation artifacts, and chirp waveforms create time-varying frequency-dependent profiles. These clear visual distinctions confirm that CSI carries rich discriminative information, yet they also reveal \textit{why building a reliable detector is non-trivial}.
 
\noindent \textbf{{The zero-day challenge (\textbf{RQ2}).}}~The combination of strategies, waveforms, and power levels produces an open-ended attack space that no finite training set can cover. A supervised classifier trained on known types will confidently assign a novel waveform to the nearest known category rather than flagging it as unseen. This closed-world assumption, inherent in any purely supervised detector, motivates the need for anomaly detection mechanisms that can score how far an input deviates from the training distribution, complementing classification with a principled \textit{``I don't know''} capability.
 
\noindent \textbf{{The adversarial challenge (\textbf{RQ3}).}}~Any detector based on machine learning is exposed to adversarial evasion, carefully crafted perturbations that cause misclassification~\cite{goodfellow2014explaining}. In the wireless domain, adversarial attacks have been demonstrated against signal classifiers~\cite{sadeghi2019adversarial}, communication decoders~\cite{bahramali2021robust}, and CSI-based sensing systems~\cite{magmaw2025, li2024practical}. Critically, perturbations transmitted over-the-air (OTA) are subject to physical constraints (e.g., power spectral density bounds, temporal correlation, subcarrier coupling) that purely digital attacks ignore~\cite{chernikova2022fence}. Evaluating a detector only against unconstrained perturbations overstates true vulnerability~\cite{athalye2018obfuscated}. Any credible robustness claim must be validated under physically realizable conditions.

\noindent \textbf{{The deployment challenge (\textbf{RQ4}).}}~IIoT environments impose a strict computational hierarchy. Field-level microcontrollers cannot run complex inference pipelines, but the supervisory level can, provided the analysis workload is reduced to only genuinely suspicious traffic. A practical detection system must decompose its computation to match this hierarchy, performing inexpensive screening at the field level and reserving detailed analysis for an edge 
node.

These three challenges, generalization, resilience, and efficiency, must be addressed jointly. A system that solves any two without the third is either \textit{fragile} (no adversarial evaluation), \textit{narrow} (no zero-day coverage), or \textit{impractical} (no edge deployment). Section~\ref{sec:design} presents how \textsc{Citadel} addresses all three through a two-stage architecture with complementary detection paradigms.


\section{Threat Model}
\label{sec:threat_model}

We consider an IIoT deployment aligned with the ISA-95/Purdue reference architecture~\cite{isa95, serror2021challenges}. Wi-Fi-enabled microcontrollers attached to industrial equipment collect telemetry and report it to the supervisory level over Wi-Fi. CSI is extracted passively from every frame exchange at no additional hardware cost, meaning physical-layer monitoring piggybacks on existing operational infrastructure. Stage~1 runs on the same microcontroller that performs the sensing task. Stage~2 runs on a GPU-capable edge node at the supervisory level, within the operational technology (OT) zone with no dependency on external cloud services, consistent with IEC~62443~\cite{iec62443}. The field-level sensors are physically accessible to an adversary (deployed on the factory floor), while the edge node resides in a secured control room.

\begin{figure}[t]
    \centering
    \includegraphics[width=\columnwidth]{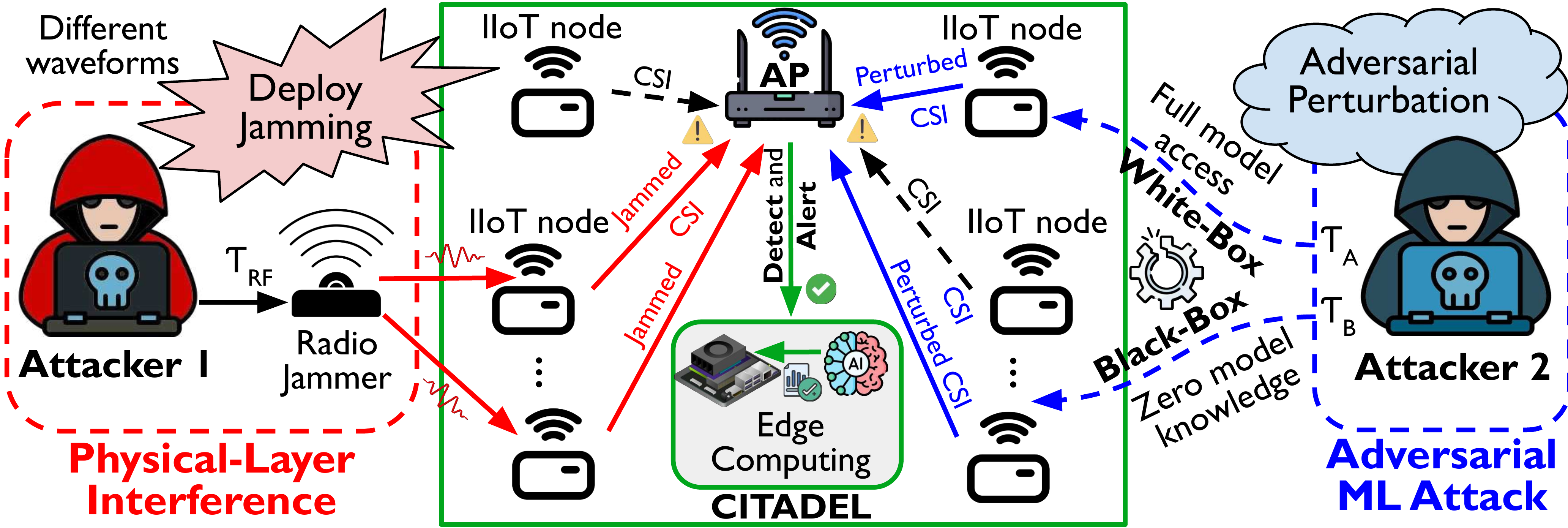}
    \caption{Threat models. The RF jammer ($\mathcal{T}_{\text{RF}}$) targets the wireless link with arbitrary waveforms. The ML attacker operates under different adversarial capabilities: white-box ($\mathcal{T}_A$) and black-box ($\mathcal{T}_B$).}
    \label{fig:threat_model}
\end{figure}

\noindent \textbf{\textit{{Attacker model.}}}~We consider two orthogonal adversary types (Figure~\ref{fig:threat_model}). An \textit{RF jammer} ($\mathcal{T}_{\text{RF}}$) deploys a SDR near the monitored link, controlling three axes: timing strategy (constant, sweeping, pulsed), waveform shape (Gaussian, QPSK, chirp, FSK, sawtooth, triangle), and transmission power~\cite{ali2022jamrf, pirayesh2022jamming}. The combination produces an open-ended attack space of which any finite training set covers only a subset. An \textit{adversarial ML attacker} crafts input perturbations to make the pipeline misclassify jammed CSI as benign. We focus on evasion attacks (hiding real jamming) rather than causative attacks (triggering false alarms), as the latter can be mitigated through alert rate limiting while missed attacks risk irreversible physical consequences. We define two threat models with progressively increasing realism, following the escalation methodology in~\cite{carlini2019evaluating, tramer2020adaptive}.

\noindent \textbf{\textit{$\mathcal{T}_A$ (\textit{White-box}):}}~The adversary has full knowledge of all model architectures and parameters, computing exact gradients. Perturbations must satisfy physical realizability constraints adapted to the CSI domain~\cite{chernikova2022fence}. 
These are implemented as a differentiable projection $\Pi_{\text{phys}}$ applied after each gradient step, enforcing: \textbf{(C1)}~per-subcarrier power within the noise floor scaled by the perturbation budget; \textbf{(C2)}~temporal coherence across consecutive time steps, reflecting the physical channel coherence time; and \textbf{(C3)}~subcarrier coupling via frequency-domain smoothing, enforcing the correlation structure imposed by indoor multipath propagation. This is our primary evaluation setting, since perturbations that violate these constraints cannot be realized over the air~\cite{chernikova2022fence, athalye2018obfuscated}.
Perturbation budgets are expressed in z-score normalized CSI space. The physical interpretation and budget selection are detailed in Section~\ref{sec:eval_adversarial}.

\noindent \textbf{\textit{$\mathcal{T}_B$ (\textit{Black-box}):}}~The adversary has zero knowledge of model architectures, parameters, or intermediate representations, and cannot compute gradients. The attacker solves the following equation using only query access to the binary detection output $y \in \{0, 1\}$:
\begin{equation}
\label{eq:tb_opt}
\boldsymbol{\delta}^{*} \approx \arg\max_{\boldsymbol{\delta}}\; \hat{\mathcal{L}}\!\bigl(y_1, \ldots, y_Q\bigr) \quad \text{s.t.}\quad \lVert\boldsymbol{\delta}\rVert_\infty \le \varepsilon,\;\; \Pi_{\text{phys}}(\boldsymbol{\delta}) = \boldsymbol{\delta}\,,
\end{equation}
where $\hat{\mathcal{L}}$ is estimated from $Q$ queries. Alternatively, the adversary trains a surrogate model on independently collected data and transfers adversarial examples crafted against it. This is the most realistic deployment scenario, reflecting an adversary who can observe whether alerts are raised but has no access to the detection pipeline internals.

\noindent \textit{Scope exclusions.}~The adversary cannot tamper with sensor hardware, compromise the authenticated field-to-supervisory channel (a man-in-the-middle variant is analyzed in Appendix~\ref{app:mitm}), or inject fabricated CSI readings. Full-band high-power jamming that destroys the link entirely is trivially detectable through link loss and falls outside intelligent evasion.

\noindent \textbf{\textit{{Defender capabilities.}}}~The defender operates the pipeline at the supervisory level with real-time access to all sensor CSI streams but possesses no \textit{a priori} knowledge of zero-day strategies. All OOD thresholds are calibrated exclusively on in-distribution data through cross-validation, no zero-day samples are used at any stage, ensuring that reported performance reflects genuine generalization.

\section{The Design of \textsc{Citadel}}
\label{sec:design}

This section presents the design of \textsc{Citadel}. 
Stage~1 runs a binary trigger on each window to screen benign traffic and suspicious windows are escalated to Stage~2, which classifies them into known jamming types or flags them as zero-day anomalies through a multi-signal ensemble. The design is guided by two principles developed throughout this section. \textit{Computational asymmetry}, which exploits the fact that the vast majority of traffic is benign to avoid expensive analysis on every window, and \textit{signal complementarity}, which ensures that no single adversarial perturbation can simultaneously suppress all detection channels.

\subsection{Stage 1: Binary Trigger}
\label{sec:stage1}

Stage~1 addresses the deployment challenge (\textbf{RQ4}) by performing inexpensive screening at the field level. For each CSI window $\mathbf{x} \in \mathbb{R}^{2 \times 32 \times 52}$, a lightweight binary classifier decides whether to escalate to Stage~2:
\begin{equation}
y_1 = \sigma\bigl(f_{S1}(\mathbf{x};\theta_1)\bigr), \quad
\text{escalate if } y_1 > \tau_1,
\label{eq:stage1}
\end{equation}
where $\sigma$ is the sigmoid function and $\tau_1$ 
is calibrated to target a low false escalation rate on benign traffic.
The architecture consists of three convolutional blocks (Conv2d, BatchNorm, ReLU) with channel progression $2 \to 4 \to 8 \to 16$, followed by adaptive average pooling and a single linear layer, yielding 1,362 trainable parameters. This extreme compactness is \textit{deliberate}. It satisfies the memory and latency constraints of microcontroller deployment while limiting the model's capacity to learn sharp decision boundaries that larger networks expose to gradient-based exploitation.

Training uses cross-entropy loss with two security-motivated modifications: \textit{asymmetric false-negative weighting} that penalizes missed attacks more heavily than false escalations, and \textit{label smoothing} that prevents overconfident predictions on boundary samples, particularly movement traffic, whose spectral characteristics partially overlap with low-power jamming. 

Beyond cost reduction (\textbf{RQ4}), the two-stage decomposition creates a structural defense against adversarial evasion (\textbf{RQ3}). An adversary must simultaneously craft perturbations that bypass Stage~1's binary filter (making jammed CSI appear benign) \textit{and} evade Stage~2's multi-signal scoring (appearing in-distribution across three independent channels). This creates a multi-objective optimization problem. As we show in Section~\ref{sec:evaluation}, the gradients of the two stages point in opposing directions, preventing simultaneous evasion.

\subsection{Stage 2: Multi-Component Analysis}
\label{sec:stage2}

\begin{figure}[t]
  \centering
  \includegraphics[width=\columnwidth]{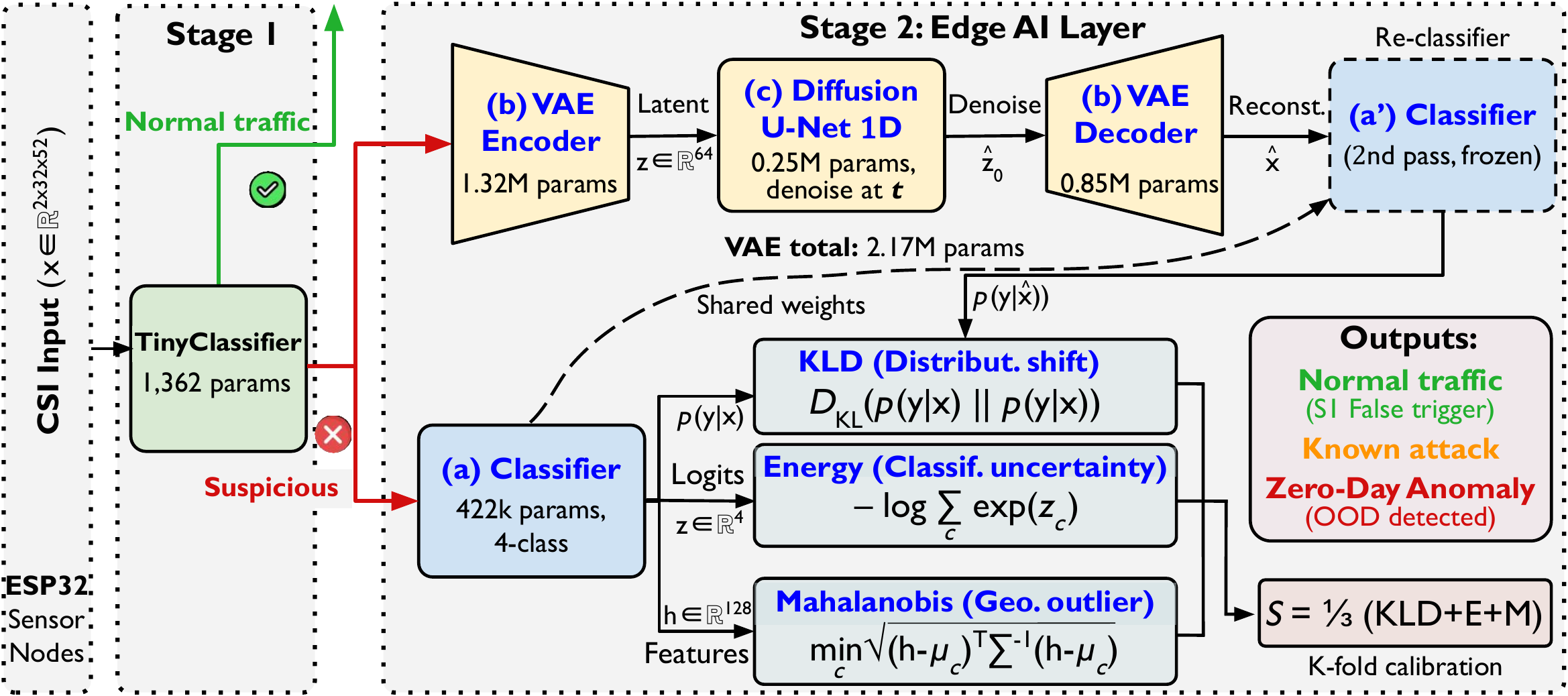}
  \caption{\textsc{Citadel} Stage~2 architecture. The CSI classifier produces logits $\mathbf{z}$ and features $\mathbf{h}$ directly from the input. The VAE encoder, diffusion denoiser, and VAE decoder reconstruct $\hat{\mathbf{x}}$, which is re-classified to obtain $p(\mathbf{y}|\hat{\mathbf{x}})$. Three OOD signals are derived from these outputs and fused via an equal-weight ensemble with K-fold calibration.}
  \label{fig:architecture}
\end{figure}

A single supervised classifier cannot satisfy the remaining two challenges. Against the zero-day challenge (\textbf{RQ2}), a classifier trained on known attack types assigns confident predictions to novel waveforms rather than flagging them as unseen, the closed-world failure described above. Against the adversarial challenge (\textbf{RQ3}), a single anomaly signal provides the adversary a clear optimization target: \textit{suppress that one signal} and the detector is evaded. \textsc{Citadel} addresses both by combining three complementary models that produce qualitatively different representations (Figure~\ref{fig:architecture}), ensuring that an attacker cannot suppress distributional shift in output space, maintain high classifier confidence, and remain geometrically close to known-class centroids in feature space simultaneously.

\subsubsection{Four-Class CSI Classifier}
\label{sec:classifier}

The backbone classifier maps CSI windows to four classes: benign~(0), constant~(1), sweeping~(2), and pulse~(3) jamming. Movement traffic is intentionally excluded from the training classes, making it OOD by design. This decision ensures the anomaly detection mechanism is continuously exercised against the natural population of movement windows during deployment, providing ongoing validation that the detector remains calibrated \textit{without} relying on attack data.

The architecture employs four convolutional layers with channel progression $32 \to 64 \to 128 \to 256$, each followed by a \textit{spectral-attention} module that implements channel recalibration~\cite{hu2018squeeze}, learning to emphasize subcarrier bands most disrupted by interference. Fully connected layers ($256 \to 128 \to 4$) with dropout produce three outputs consumed by downstream components: the class prediction $\hat{c} = \arg\max_c z_c$, the logit vector $\mathbf{z} \in \mathbb{R}^4$ for energy-based scoring, and the penultimate feature representation $\mathbf{h} \in \mathbb{R}^{128}$ for distance-based scoring. The classifier totals 422,500 parameters.

\subsubsection{Variational Autoencoder}
\label{sec:vae}

A CNN-based VAE~\cite{kingma2014vae} with spectral-attention modules learns a generative model of the in-distribution CSI manifold through a 64-dimensional latent space. The encoder mirrors the classifier's convolutional structure but maps to separate mean ($\boldsymbol{\mu}$) and log-variance ($\log\boldsymbol{\sigma}^2$) heads. The decoder uses transposed convolutions to reconstruct the original $2 \times 32 \times 52$ tensor, totaling 2.17M parameters. 
The training loss combines three terms:
\begin{equation}
\mathcal{L}_{\text{VAE}} = \mathcal{L}_{\text{recon}} +
\lambda_{\text{KL}} \cdot D_{\text{KL}}\bigl(q(\mathbf{z}|\mathbf{x}) \,\|\,
p(\mathbf{z})\bigr) + \lambda_{\text{perc}} \cdot \mathcal{L}_{\text{perc}},
\label{eq:vae_loss}
\end{equation}
where $\mathcal{L}_{\text{recon}}$ is the mean squared error (MSE) reconstruction loss, $D_{\text{KL}}$ regularizes against the standard normal prior $p(\mathbf{z}) = \mathcal{N}(\mathbf{0}, \mathbf{I})$, and $\mathcal{L}_{\text{perc}}$ is a perceptual loss in classifier feature space that encourages reconstructions to preserve semantically meaningful structure.

\noindent \textbf{{Distribution-shift adaptation.}}~During training, the VAE sees all benign windows. During deployment, Stage~2 receives only the Stage~1-escalated subset, which is enriched in samples with unusual spectral characteristics. Without adaptation, this distribution shift inflates Stage~2 false alarms. We address this through fine-tuning on Stage~1-gated benign samples with a replay buffer to prevent catastrophic forgetting, closing the gap at negligible training cost.

\subsubsection{Latent-Space Diffusion Model}
\label{sec:diffusion}

A 1D U-Net~\cite{ronneberger2015unet} with sinusoidal time embeddings and residual blocks is trained as a denoising diffusion probabilistic model (DDPM)~\cite{ho2020ddpm} on the VAE's latent representations $\mathbf{z} \in \mathbb{R}^{64}$, totaling 249,600 parameters. Operating in the 64-dimensional latent space rather than the 3,328-dimensional input space is critical for computational feasibility on edge hardware and concentrates the model's capacity on the manifold structure learned by the VAE.
At inference, single-step denoising produces a denoised latent $\hat{\mathbf{z}}_0$. The VAE decoder maps this back to input space as $\hat{\mathbf{x}}$, and the classifier's output distributions on the original versus reconstructed input are compared via KL divergence:
\begin{equation}
\text{KLD}(\mathbf{x}) = D_{\text{KL}}\bigl(
p(\mathbf{y}|\mathbf{x}) \,\|\, p(\mathbf{y}|\hat{\mathbf{x}})
\bigr).
\label{eq:kld}
\end{equation}
For in-distribution inputs, the VAE latent lies on the learned manifold. The denoise cycle preserves the classifier's output distribution, yielding low KLD. For OOD inputs, zero-day attacks or adversarially perturbed windows, the latent lies off the manifold, and denoising pulls it toward the nearest in-distribution configuration, distorting the output distribution and producing elevated KLD. This mechanism converts high-dimensional feature anomalies into a measurable divergence in the compact 4-class output space, amplifying subtle distributional shifts that raw feature-space metrics might miss. Notably, the diffusion model also provides implicit robustness. Its stochastic denoising acts as a form of randomized smoothing~\cite{nie2022diffpure}, making the KLD signal resistant to \textit{gradient-based manipulation}.

\subsection{Ensemble OOD Detection}
\label{sec:ensemble}

\textsc{Citadel} fuses three anomaly signals, each capturing a qualitatively different aspect of distributional deviation. Their combination forces an adversary to simultaneously suppress distributional shift in output space, maintain high classifier confidence, and remain geometrically close to known-class centroids in feature space, a multi-objective constraint that no single perturbation direction can satisfy.

\noindent \textbf{{Signal 1: KL Divergence.}}~The distributional shift between classifier outputs on original versus diffusion-reconstructed inputs (Equation~\ref{eq:kld}). This signal captures how fragile the classifier's confidence allocation is under manifold-constrained reconstruction.

\noindent \textbf{{Signal 2: Energy Score.}}~The negative log-sum-exp of the logit vector~\cite{liu2020energy}:
\begin{equation}
E(\mathbf{x}) = -\log \sum_{c} \exp(z_c / T),
\label{eq:energy}
\end{equation}
where temperature $T$ sharpens the energy landscape. Higher energy indicates lower model confidence. OOD inputs produce elevated energy because learned features fail to activate any class prototype strongly.

\noindent \textbf{Signal 3: Mahalanobis Distance.}~The minimum distance from class centroids in the classifier's penultimate feature space~\cite{lee2018mahalanobis}:
\begin{equation}
M(\mathbf{x}) = \min_{c} \sqrt{
(\mathbf{h} - \boldsymbol{\mu}_c)^\top
\boldsymbol{\Sigma}^{-1}
(\mathbf{h} - \boldsymbol{\mu}_c)
},
\label{eq:mahalanobis}
\end{equation}
where $\boldsymbol{\mu}_c \in \mathbb{R}^{128}$ are class-conditional centroids and $\boldsymbol{\Sigma}^{-1}$ is the shared precision matrix estimated via Ledoit-Wolf shrinkage~\cite{ledoit2004well}. This signal flags inputs in low-density regions of the feature manifold, even when the classifier assigns high softmax confidence through extrapolation.

Each signal is normalized to $[0,1]$ via soft thresholding against calibration bounds, and the ensemble score is the unweighted mean:
\begin{equation}
S_{\text{ens}} = \tfrac{1}{3}\, s_{\text{KLD}} + \tfrac{1}{3}\, s_{\text{energy}}
+ \tfrac{1}{3}\, s_{\text{Mahal}}.
\label{eq:ensemble}
\end{equation}
A sample is flagged as OOD if $S_{\text{ens}}$ exceeds a class-conditional threshold or the raw KLD exceeds an extreme-override percentile. Equal weighting is a deliberate design choice. Optimized weights would give an adversary a clear target (suppress the dominant signal), whereas equal weights require suppressing all three simultaneously.

\textbf{\textit{K-fold calibration.}}~All OOD thresholds are set through 5-fold blocked temporal cross-validation with purge gaps that eliminate both double-dipping bias and temporal leakage from CSI autocorrelation. The procedure operates exclusively on in-distribution data. No zero-day samples are used at any stage, ensuring that reported zero-day performance reflects genuine generalization rather than implicit tuning. Class-conditional thresholds use a lower percentile for benign-predicted samples (reducing false positives on benign traffic) and a higher percentile for attack-predicted samples (ensuring that OOD signals on putative attacks are taken seriously).

\section{Implementation and Setup}
\label{sec:implementation}

\begin{figure}[t]
\centering
\includegraphics[width=\columnwidth]{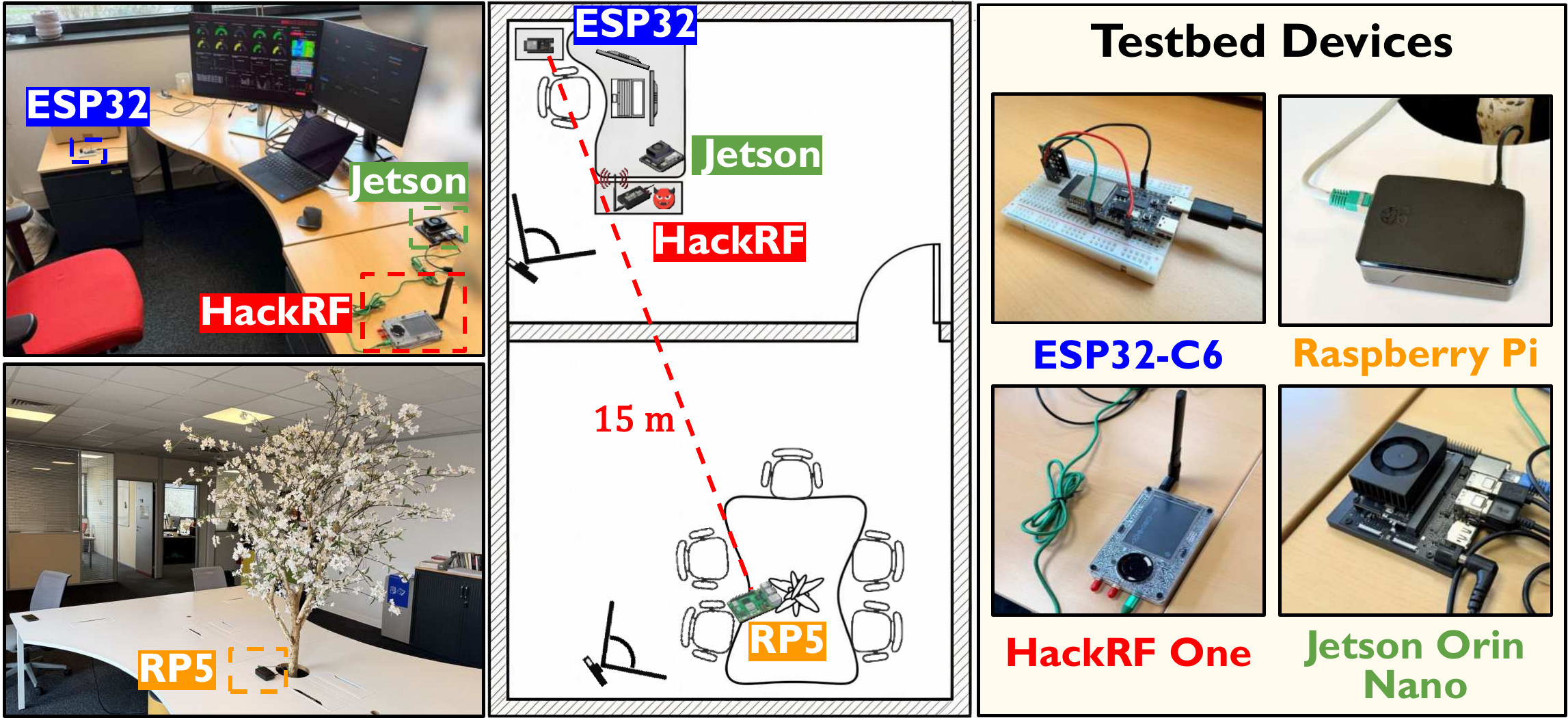}
\caption{Experimental testbed. (Left)~Laboratory environment. (Middle)~Floor plan showing ESP32-C6 nodes, Jetson Orin Nano, HackRF One, and Raspberry Pi~5 access point. (Right)~Hardware components.}
\label{fig:testbed}
\end{figure}

\noindent \textbf{Testbed.}~The experimental testbed (Figure~\ref{fig:testbed}) emulates an IIoT factory floor. Five ESP32-C6 microcontrollers~\cite{Espressif_CSI_2025} serve as Wi-Fi-enabled sensor nodes, extracting per-subcarrier CSI from 802.11n frames on the 2.4\,GHz band (20\,MHz channel, 52 OFDM subcarriers) at approximately 4\,Hz while performing their nominal sensing tasks. A Raspberry Pi~5 acts as both access point and MQTT broker. An NVIDIA Jetson Orin Nano at the supervisory level executes Stage~2 inference. A HackRF One SDR~\cite{ali2022jamrf, portapack_mayhem} generates jamming signals across three timing strategies (constant, sweeping, pulsed), six waveform types (Gaussian, QPSK, chirp, FSK, sawtooth, triangle), and three power levels (10, 15, 20\,dB IF gain). The jammer is placed at variable distances in a $4 \times 5$\,m room, with the access point 15\,m away to introduce realistic multipath conditions.

\begin{table}[t]
\scriptsize
\centering
\caption{Training configuration by pipeline component.}
\label{tab:training}
\resizebox{\columnwidth}{!}{%
\begin{tabular}{lcccl}
\toprule
\textbf{Component} & \textbf{Epochs} & \textbf{LR} & \textbf{Batch} & \textbf{Loss / Notes} \\
\midrule
S1 TinyClassifier & 100 & $10^{-3}$ & 256 & CE; label smooth $\alpha{=}0.1$; FN $2{\times}$ \\
S2 CSI Classifier & 50 & $10^{-3}$ & 128 & CE; $L_2$ decay $10^{-4}$ \\
S2 VAE & 100 & $10^{-4}$ & 256 & Eq.~\ref{eq:vae_loss}; $\lambda_{\text{KL}}{=}10^{-6}$ \\
\quad + Fine-tune & 30 & $2{\times}10^{-5}$ & 256 & S1-gated quiet; 30\% replay \\
S2 Diffusion U-Net & 1000 & $10^{-5}$ & 1024 & DDPM; EMA 0.9999 \\
\bottomrule
\end{tabular}}
\end{table}

\noindent \textbf{Data collection.}~Each ESP32-C6 transmits raw CSI packets ($I_k$ and $Q_k$ for 52 subcarriers) via MQTT. The preprocessing pipeline converts these to model-ready tensors: \tikz[baseline=(char.base)]\node[shape=circle, fill=black, inner sep=0.7pt, text=white] (char) {1};~I/Q pairs yield complex estimates $H_k = I_k + jQ_k$; \tikz[baseline=(char.base)]\node[shape=circle, fill=black, inner sep=0.7pt, text=white] (char) {2};~magnitude and phase are computed per subcarrier; \tikz[baseline=(char.base)]\node[shape=circle, fill=black, inner sep=0.7pt, text=white] (char) {3};~corrupt packets (flagged by the firmware) are filtered; \tikz[baseline=(char.base)]\node[shape=circle, fill=black, inner sep=0.7pt, text=white] (char) {4};~the stream is segmented into sliding windows of $W{=}32$ consecutive packets, yielding tensors $\mathbf{x} \in \mathbb{R}^{2 \times 32 \times 52}$. The dataset comprises over 815,000 labeled (Appendix~\ref{app:dataset}) windows: 264,014 benign (quiet and movement), 366,796 from 18 known-attack scenarios, and 184,508 from 15 zero-day scenarios strictly withheld (Appendix~\ref{app:coverage} details the coverage matrix) from all training, validation, and calibration. Data is partitioned temporally (70/15/15 train/val/test) with 128-window purge gaps to eliminate autocorrelation-driven leakage.

\begin{table*}[t]
\centering
\caption{\textsc{Citadel} zero-day detection across 15 held-out scenarios. All values are percentages of total windows per scenario. Stage~2 decisions: OOD = correctly flagged as unknown threat; Constant/Sweeping/Pulse = absorbed into a known class (still detected, not a failure); Benign = missed by the full pipeline. Left: Category~1 (novel timing). Right: Category~2 (known timing, novel waveform).}
\label{tab:zeroday}
\scriptsize
\setlength{\tabcolsep}{4pt}
\begin{tabular}{ll r r r r r r !{\vrule width 1.2pt} ll r r r r r r}
\toprule
& & \textbf{S1}
& \multicolumn{5}{c}{\textbf{Stage~2 Decision (\%)}}
& & & \textbf{S1}
& \multicolumn{5}{c}{\textbf{Stage~2 Decision (\%)}} \\
\cmidrule(lr){3-3}\cmidrule(lr){4-8}
\cmidrule(lr){11-11}\cmidrule(l){12-16}
\textbf{Pattern}
  & \textbf{Waveform}
  & \textbf{Trig.}
  & \textbf{Ben.}
  & \textbf{Con.}
  & \textbf{Swe.}
  & \textbf{Pul.}
  & \textbf{OOD}
  & \textbf{Pattern}
  & \textbf{Waveform}
  & \textbf{Trig.}
  & \textbf{Ben.}
  & \textbf{Con.}
  & \textbf{Swe.}
  & \textbf{Pul.}
  & \textbf{OOD} \\
\midrule
\multicolumn{8}{l}{\textit{Category 1: Novel timing}}
  & \multicolumn{8}{l}{\textit{Category 2: Known timing, novel waveform}} \\
\addlinespace[2pt]
Random & Bruteforce & 100.0 & 0.0 & 0.0 & 0.0  & 0.0 & \textbf{100.0}
  & Constant & Chirp      & 100.0 & 0.0 & 0.0 & 25.2 & 0.0 & \textbf{74.8} \\
Random & Chirp      & 100.0 & 0.0 & 0.0 & 0.0  & 0.0 & \textbf{100.0}
  & Sweeping & Bruteforce & 100.0 & 0.0 & 0.0 & 0.2  & 0.0 & \textbf{99.8} \\
Random & FSK        & 100.0 & 0.0 & 0.0 & 0.0  & 0.0 & \textbf{100.0}
  & Sweeping & FSK        & 100.0 & 0.0 & 0.0 & 0.0  & 0.0 & \textbf{100.0} \\
Random & Sawtooth   & 100.0 & 0.0 & 0.0 & 0.0  & 0.0 & \textbf{100.0}
  & Sweeping & Sawtooth   & 100.0 & 0.0 & 0.0 & 0.0  & 0.0 & \textbf{100.0} \\
Random & Square     & 100.0 & 0.0 & 0.0 & 0.0  & 0.0 & \textbf{100.0}
  & Pulse    & Bruteforce & 100.0 & 0.0 & 0.0 & 4.9  & 0.0 & \textbf{95.1} \\
Random & Triangle   & 100.0 & 0.0 & 0.0 & 0.0  & 0.0 & \textbf{100.0}
  & Pulse    & Chirp      & 100.0 & 0.0 & 0.0 & 0.0  & 0.0 & \textbf{100.0} \\
Burst  & Chirp      & 100.0 & 0.1 & 0.2 & 0.0  & 0.0 & \textbf{99.7}
  & Pulse    & Triangle   & 100.0 & 0.0 & 0.0 & 0.0  & 0.0 & \textbf{100.0} \\
Burst  & Triangle   &  97.5 & 5.0 & 0.0 & 7.5  & 0.0 & \textbf{87.5}
  & & & & & & & \\
\cmidrule(lr){3-8}\cmidrule(lr){11-16}
\multicolumn{2}{r}{\textit{Mean}}
  & \textit{99.7} & & & & & \textit{98.4}
  & \multicolumn{2}{r}{\textit{Mean}}
  & \textit{100.0} & & & & & \textit{95.7} \\
\midrule
\multicolumn{16}{c}{%
  \textbf{Overall mean (15 scenarios):}\quad
  Stage~1 Trigger = \textbf{99.8\%}\,,\quad
  OOD (anomaly detection) = \textbf{97.1\%}} \\
\bottomrule
\end{tabular}
\end{table*}

\noindent \textbf{Training.}~All models are trained sequentially on a single NVIDIA L40S GPU. Table~\ref{tab:training} lists the complete configuration (convergence curves are provided in Appendix~\ref{app:training}). The Stage~1 threshold $\tau_1$ targets less than 10\% false escalation. The VAE's KL weight is $\lambda_{\text{KL}}{=}10^{-6}$.  The fine-tuning phase described in Section~\ref{sec:vae} closes the Stage~1-gating distribution gap. The diffusion model uses 1,000 DDPM timesteps ($\beta$ from $10^{-4}$ to $0.02$). At inference, denoising at $t{=}10$ balances anomaly sensitivity against reconstruction fidelity. For ensemble calibration (Section~\ref{sec:ensemble}), 5-fold blocked temporal CV sets per-signal bounds at P95 (lower) and in-distribution maximum (upper), with KLD extreme override at P99.5. 
Class-conditional ensemble thresholds are set at P90 for benign-predicted and P95 for attack-predicted samples, and energy temperature is $T{=}0.5$. Full calibration parameters are listed in Appendix~\ref{app:calibration}.

\noindent \textbf{Evaluation metrics.}~We report six metrics throughout. On the \textit{defense} side: (1)~\textit{Detection rate (DR)}: fraction of attack windows correctly identified as threats, combining Stage~1 and Stage~2 decisions; (2)~\textit{Stage~1 trigger rate}: fraction of windows flagged as suspicious by the binary trigger, determining what reaches Stage~2; (3)~\textit{OOD detection rate}: fraction of windows flagged as out-of-distribution by the ensemble, the primary metric for zero-day scenarios; (4)~\textit{False positive rate (FPR)}: fraction of benign windows incorrectly flagged as threats. On the \textit{adversarial} side: (5)~\textit{Evasion rate (ER)}: fraction of attack windows that evade detection, reported at Stage~1, Stage~2, and end-to-end (E2E) levels to isolate each component's contribution; (6)~\textit{Attack success rate (ASR)}: fraction of adversarial examples that evade the targeted component or pipeline.

\section{Evaluation}
\label{sec:evaluation}

We evaluate \textsc{Citadel} along three dimensions: \tikz[baseline=(char.base)]\node[shape=circle,draw,inner sep=0.7pt] (char) {1};~detection performance on known and zero-day attacks under non-adversarial conditions (Section~\ref{sec:eval_detection}); \tikz[baseline=(char.base)]\node[shape=circle,draw,inner sep=0.7pt] (char) {2};~adversarial robustness under the threat models defined in Section~\ref{sec:threat_model} (Section~\ref{sec:eval_adversarial}); \tikz[baseline=(char.base)]\node[shape=circle,draw,inner sep=0.7pt] (char) {3};~comparative evaluation against eight baseline methods, including inference efficiency on edge hardware and operational assessment (Section~\ref{sec:eval_comparative}). All results use the testbed and training configuration described in Section~\ref{sec:implementation}.

\subsection{Detection Performance}
\label{sec:eval_detection}

Under non-adversarial conditions, \textsc{Citadel} achieves 100\% known-attack detection, 97.1\% zero-day anomaly detection across 15~held-out scenarios, and 0.4\% E2E FPR (\textbf{RQ1}, \textbf{RQ2}). The 100\% known-attack rate establishes a critical baseline for interpreting the adversarial results in Section~\ref{sec:eval_adversarial}. Any reduction under perturbation reflects genuine adversarial degradation rather than residual clean-condition error. We now examine each dimension in detail.

\begin{table}[t]
\centering
\caption{\textsc{Citadel} two-stage known-attack detection. Each known class contains Gaussian and QPSK waveforms at three power levels (10, 15, 20\,dB; 6 scenarios per class, 18 total). Stage~2 decisions are shown as percentages of all windows per class.}
\label{tab:known}
\scriptsize
\adjustbox{max width=\columnwidth}{%
\begin{tabular}{l r r r r r r}
\toprule
& \textbf{Stage~1}
& \multicolumn{5}{c}{\textbf{Stage~2 Decision (\%)}} \\
\cmidrule(lr){2-2}\cmidrule(l){3-7}
\textbf{Class}
  & \textbf{Trigger (\%)}
  & \textbf{Benign}
  & \textbf{Constant}
  & \textbf{Sweeping}
  & \textbf{Pulse}
  & \textbf{OOD} \\
\midrule
Constant & 100.0              & 0.0           & \textbf{96.8} & 0.2           & 0.0            & 3.0  \\
Sweeping & 100.0              & 0.0           & 0.0           & \textbf{83.1} & 0.0            & 16.9 \\
Pulse    & 100.0              & 0.0           & 0.0           & 0.0           & \textbf{100.0} & 0.0  \\
\addlinespace
Benign   & \textit{8.6 (FPR)} &               &               &               &                & \textbf{E2E: 0.4\%} \\
\bottomrule
\end{tabular}}
\end{table}

\subsubsection{Known-Attack Classification}
\label{sec:eval_known}

Table~\ref{tab:known} disaggregates detection performance across the two pipeline stages for each known attack class and benign traffic. Stage~1 achieves 100\% trigger rate on all three attack classes, meaning every jamming window is escalated to the supervisory level for detailed analysis, and the 8.6\% benign trigger rate (FPR) remains below the 10\% operational target, with the majority of false triggers originating from movement-induced CSI fluctuations that partially overlap the spectral profile of low-power jamming.

Stage~2 correctly classifies the vast majority of escalated known-attack windows into their true class. Pulse jamming is classified with perfect accuracy (100.0\%), constant jamming at 96.8\%, and sweeping jamming at 83.1\%. The remainder is flagged as OOD rather than misclassified into a wrong attack class, a conservative failure mode that preserves detection (the window is still recognized as a threat) while sacrificing root-cause specificity. This mischaracterization is concentrated in sweeping attacks (16.9\% routed to OOD), where the time-varying frequency pattern of QPSK waveforms at stealthy power levels produces spectral profiles that fall between training-class centroids in the classifier's feature space. Crucially, no known-attack window is misclassified as benign. The 0.0\% benign column across all three attack classes means that the combination of supervised classification and OOD detection achieves perfect threat identification even when class assignment is uncertain. The E2E FPR of 0.4\% confirms that the cascaded two-stage architecture suppresses false alarms. 
From the 8.6\% of benign windows that trigger Stage~1, Stage~2 correctly identifies the vast majority as non-threatening, yielding a final false-alarm rate well below the 5\% operational target.

\subsubsection{Zero-Day Generalization}
\label{sec:eval_zeroday}

Table~\ref{tab:zeroday} presents the per-scenario breakdown across 15 held-out zero-day scenarios that were \textit{strictly withheld} from all training, validation, and calibration stages. The scenarios are organized into two novelty categories: category~1 (novel timing pattern with arbitrary waveform, 8 scenarios) tests whether the OOD ensemble can detect timing structures absent from training; category~2 (known timing with novel waveform, 7 scenarios) tests whether it can detect waveforms whose spectral profiles were never seen, even when the timing pattern matches a training class.
All six random-family scenarios achieve perfect OOD detection. The random timing pattern produces broadband spectral disruption across all subcarriers simultaneously, generating KLD values well above the extreme-override threshold. The VAE cannot reconstruct a pattern that bears no structural resemblance to any training-set class. This constitutes the ``easy'' regime for the ensemble: \textit{when an attack is spectrotemporally distant from all known classes, even a single OOD signal suffices}.

Detection difficulty increases with spectral overlap between the novel waveform and training classes. The hardest category~1 scenario is \textit{burst triangle} (87.5\% OOD), where the intermittent timing reduces the observation window for distributional scoring, and the 97.5\% Stage~1 trigger rate means that 2.5\% of windows are not escalated at all. The 5.0\% benign rate confirms that a small fraction of burst windows appear sufficiently ``quiet'' to pass Stage~2 undetected, the only scenario where the pipeline shows measurable leakage. In category~2, \textit{constant chirp} is the hardest scenario (74.8\% OOD), where the chirp waveform's broadband energy partially mimics the \textit{constant gaussian} spectral profile seen during training. The remaining 25.2\% are absorbed into the sweeping class (a conservative mischaracterization that still counts as threat detection). Category~1 scenarios average 98.4\% OOD versus 95.7\% for category~2, suggesting that \textit{novel timing patterns are marginally easier to detect than novel waveforms} under known timing. This is consistent with the intuition that the classifier's learned timing features provide a strong prior that novel timing immediately violates, whereas novel waveforms may partially activate existing spectral filters.

\subsubsection{OOD Signal Analysis}
\label{sec:eval_signals}

\begin{figure}[t]
\centering
\includegraphics[width=\columnwidth]{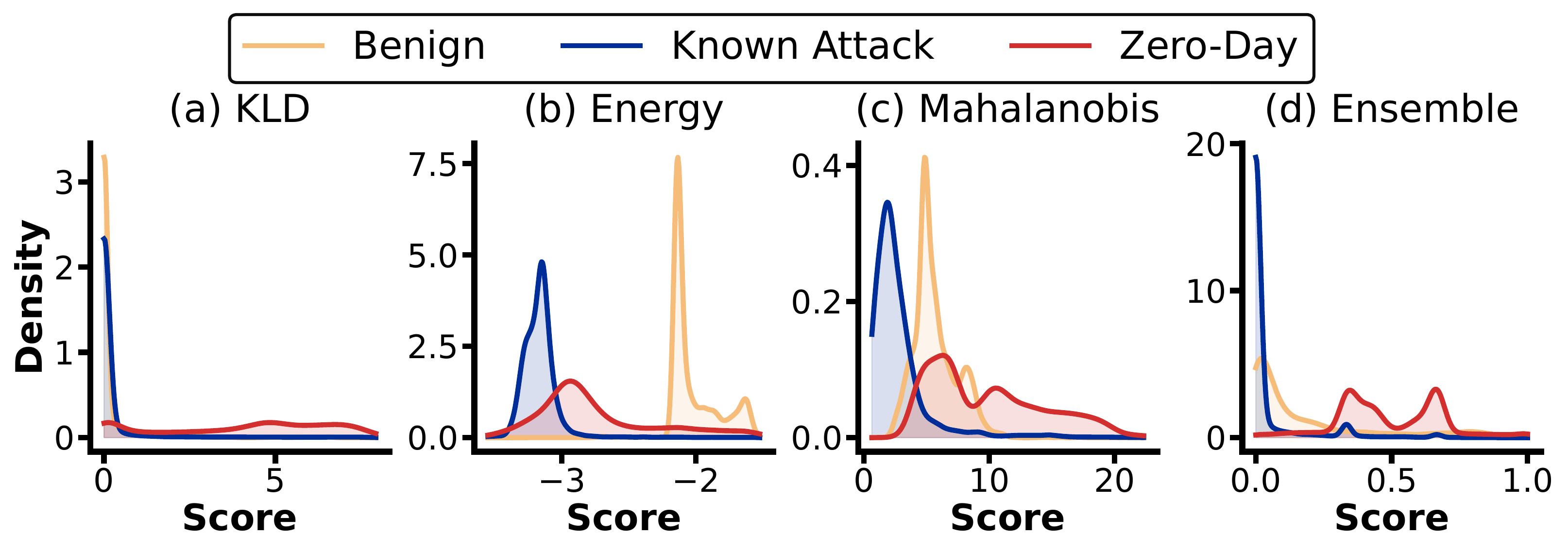}
\caption{Distribution of the three OOD signals and their ensemble fusion across benign, known-attack, and zero-day traffic (Stage~1-triggered samples). (a)--(c) show that no individual signal cleanly separates all three classes; (d) 
shows that the equal-weight ensemble achieves substantially better separation.}


\label{fig:signal_decomp}
\end{figure}

To understand \textit{why} the ensemble succeeds where individual signals fail, Figure~\ref{fig:signal_decomp} visualizes the distributions of all three OOD signals and their fusion across three traffic classes (benign, known attack, zero-day), computed on Stage~1-triggered samples only.

\noindent \textbf{KLD} (Figure~\ref{fig:signal_decomp}a) is the strongest single discriminator. Benign and known-attack samples both cluster tightly near zero, confirming that the VAE faithfully reconstructs in-distribution patterns and that diffusion denoising preserves the classifier's output distribution. Zero-day samples shift to a median of 4.57 with a wide spread (P5--P95: 0.00--7.39), providing 20 times separation from the known-attack median. Over 80\% of zero-day samples exceed  a KLD of 1.0 and 67\% exceed a KLD of 3.0, triggering the extreme-override threshold. However, 11.6\% of zero-day samples have a KLD below 0.1. The VAE partially reconstructs waveforms that are spectrally similar to training classes (e.g., constant\_chirp). This blind spot is precisely where the other two signals \textit{must compensate}.

\noindent \textbf{Energy} (Figure~\ref{fig:signal_decomp}b) provides the classifier-confidence dimension. Known attacks produce the most confident predictions (median $-3.17$), reflecting strong alignment with learned class prototypes. Benign triggered samples are less confident (median $-2.12$), and zero-day samples sit in between (median $-2.88$) with a bimodal distribution. Some novel waveforms produce confident misclassifications (the classifier believes they belong to a known class), while others genuinely confuse the classifier. Energy alone achieves only 8.4\% zero-day anomaly, 
confirming that classifier confidence is a weak OOD signal in the CSI domain. Nevertheless, energy contributes uniquely to the ensemble by providing the benign/known-attack separation that KLD and Mahalanobis \textit{lack}. In Figure~\ref{fig:signal_decomp}a and Figure~\ref{fig:signal_decomp}c, benign and known-attack distributions overlap substantially, but in Figure~\ref{fig:signal_decomp}b they are clearly \textit{distinct}.

\noindent \textbf{Mahalanobis distance} (Figure~\ref{fig:signal_decomp}c) measures geometric deviation in feature space. Known-attack samples sit closest to their respective class centroids (median 2.15), benign samples are intermediate (median 5.24), and zero-day samples are farthest (median 9.10) with a long tail reaching beyond 20. However, the overlap between benign and zero-day distributions is \textit{significant}: 49\% of zero-day samples fall below the benign P95 of 8.82. This limits standalone detection to 57.2\%, but the 51\% of zero-day samples that are above the benign range include many that KLD misses, providing \textit{complementary geometric coverage}.

\noindent \textbf{Ensemble} (Figure~\ref{fig:signal_decomp}d) demonstrates the \textit{fusion payoff}. The equal-weight combination compresses known-attack and benign scores tightly near \textit{zero} (medians 0.00 and 0.06 respectively), while shifting zero-day scores to a median of 0.45 with a broad right tail. The overlap region between benign and zero-day distributions shrinks substantially compared to any individual panel, visually confirming that the three signals cover each other's \textit{blind spots}. The low pairwise correlations between signals
confirm that each signal captures non-redundant information. Even the highest correlation (energy versus Mahalanobis, which share the same classifier backbone) leaves 82\% of variance unshared (Appendix~\ref{app:signal_correlation}).

\subsubsection{Component Ablation}
\label{sec:eval_ablation}


Adding energy to Mahalanobis yields only $+$0.6\,pp (57.8\%), confirming that without KLD, the discriminative features are fundamentally limited. KLD alone achieves 94.1\%, by far the strongest single signal, but at the cost of 9.2\% mischaracterization of known attacks, because some known-attack windows produce moderate KLD when the VAE reconstruction introduces minor distributional shifts. The full three-signal ensemble reaches 97.1\% zero-day anomaly while reducing mischaracterization from 9.2\% to 6.6\%, achieving a better detection-precision \textit{trade-off} than any single signal.


The diffusion model's marginal contribution is $+$39.0\,pp. Removing it drops detection from 96.9\% to the energy-plus-Mahalanobis baseline of 57.8\%. This confirms that the \textit{generative reconstruction channel} is the primary differentiator between \textsc{Citadel} and methods that rely solely on discriminative or distance-based scoring. This explains why post-hoc OOD methods applied to the same classifier backbone (Section~\ref{sec:eval_comparative}) achieve substantially lower zero-day rates.


\noindent \textbf{\textit{Weight sensitivity.}}~To validate the equal-weight design choice, we conduct a systematic grid search over 66 weight configurations across the 3-simplex (step size 0.1). Equal weighting achieves the highest zero-day detection rate at the P90 operating point (99.0\%). The top~15 configurations span less than 1\,pp in zero-day rate, revealing a flat optimum: \textit{the system is insensitive to moderate weight perturbations}. Skewing weights toward any single signal strictly degrades performance. Equal weighting is therefore not a simplifying assumption but the empirically justified choice, consistent with the low pairwise correlations reported in Section~\ref{sec:eval_signals}.

\noindent \textbf{\textit{Stage~1 gate ablation.}}~Forwarding all traffic directly to Stage~2 (bypassing the gate) changes zero-day anomaly below 0.1\,pp and known-attack detection remains at 100\%, confirming that Stage~2 independently distinguishes benign from malicious traffic. Stage~1's contribution is therefore \textit{computational}, not \textit{discriminative}. It filters approximately 92\% of benign windows, reducing the Stage~2 analysis volume and the associated latency and energy cost by an order of magnitude.


\subsection{Adversarial Robustness}
\label{sec:eval_adversarial}


\subsubsection{White-Box Attacks ($\mathcal{T}_A$)}
\label{sec:eval_wb}
Perturbations are bounded in the z-score normalized CSI space. For each subcarrier, $|\delta_k| \le \varepsilon$, equivalently $\varepsilon \cdot \sigma_k$ in physical units, where $\sigma_k$ is the per-subcarrier standard deviation of the training distribution. At $\varepsilon{=}0.10$, the adversary may shift each subcarrier by at most 1\,dB in magnitude and 10$^{\circ}$ in phase. We use this as the largest budget because beyond it, perturbations exceed the observed channel variability, making the adversarial manipulation distinguishable from legitimate channel dynamics.
White-box access gives the adversary exact gradients through every component in the pipeline. We follow the adaptive-attack methodology~\cite{tramer2020adaptive}: all losses are fully \textit{differentiable} (no gradient masking), the attacker optimizes directly against each target component, and we verify convergence by running PGD at 10 and 100 steps. We use PGD-100~\cite{madry2018towards} as the primary attack under $\mathcal{T}_A$ with physically realizable perturbations at $\varepsilon \in \{0.01, 0.03, 0.05, 0.10\}$. FGSM~\cite{goodfellow2014explaining} and PGD-10 converge to the same evasion rates at each $\varepsilon$, confirming that the optimization landscape is well-characterized and additional iterations do not uncover stronger attacks (\textbf{RQ3}).

\begin{figure}[t]
  \centering
  \includegraphics[width=\columnwidth]{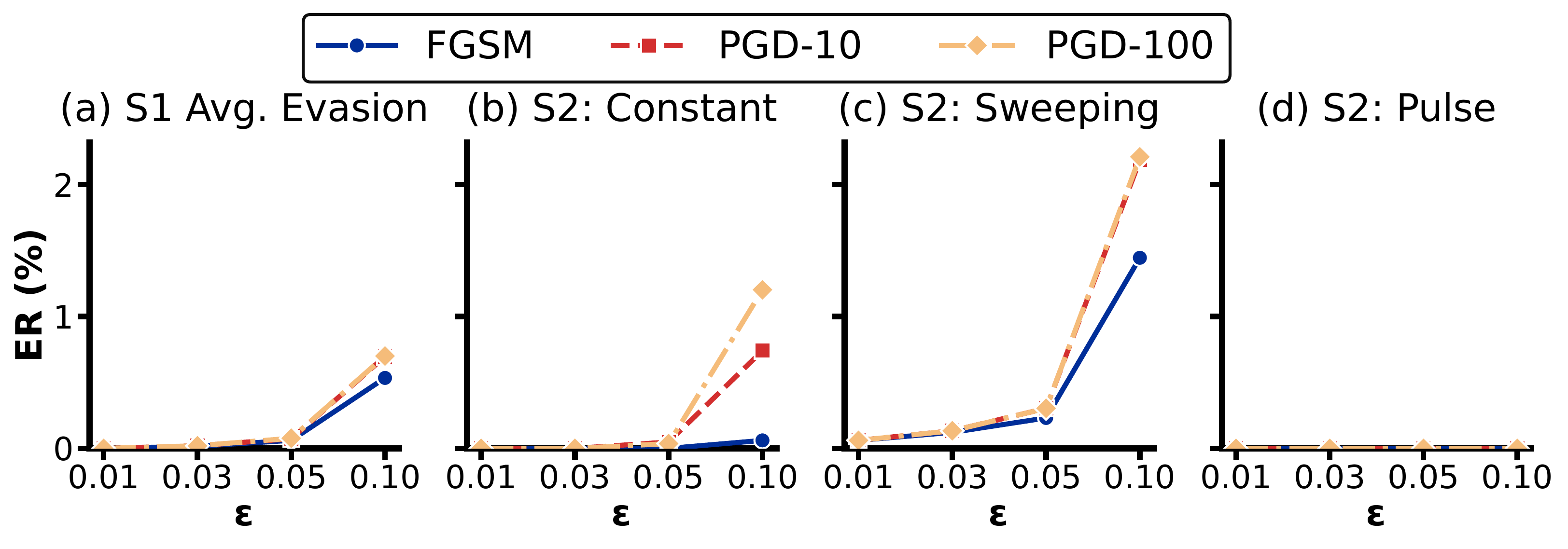}
 \caption{White-box component resilience under FGSM/PGD across perturbation budgets~$\varepsilon$. (a)~Stage~1 evasion rate averaged over known attacks;
 (b--d)~Stage~2 classifier-to-benign rate for constant, sweeping, and pulse respectively. All ERs remain below 2.3\% even at $\varepsilon{=}0.10$.}

  \label{fig:wb_component}
\end{figure}

\noindent \textbf{\textit{Stage~1 and classifier resilience.}}~Figure~\ref{fig:wb_component} isolates each detection component under targeted gradient attack. Panel~(a) averages Stage~1 evasion across all three known-attack categories. The binary trigger stays below 0.7\% bypass at $\varepsilon{=}0.10$ under PGD-100. The trigger detects broadband spectral energy as a statistical aggregate over the full 52-subcarrier band. Suppressing this aggregate while still jamming requires perturbation magnitudes that \textit{exceed} physically realizable bounds. Panels~(b--d) report the classifier's targeted-to-benign rate, where the adversary maximizes $P(\text{benign} \mid x_{\text{adv}})$. The three attack categories respond differently. Pulse jamming (panel~d) yields 0.0\% benign predictions at every $\varepsilon$. The on-off temporal structure sits too far from the benign decision region for any feasible perturbation to close the gap. Constant jamming (panel~b) reaches 1.2\%, and sweeping (panel~c) reaches 2.2\% at $\varepsilon{=}0.10$. Even under the \textit{strongest} attack configuration, fewer than 2.2\% of windows receive a benign label from the classifier.

\begin{figure}[t]
  \centering
  \includegraphics[width=\columnwidth]{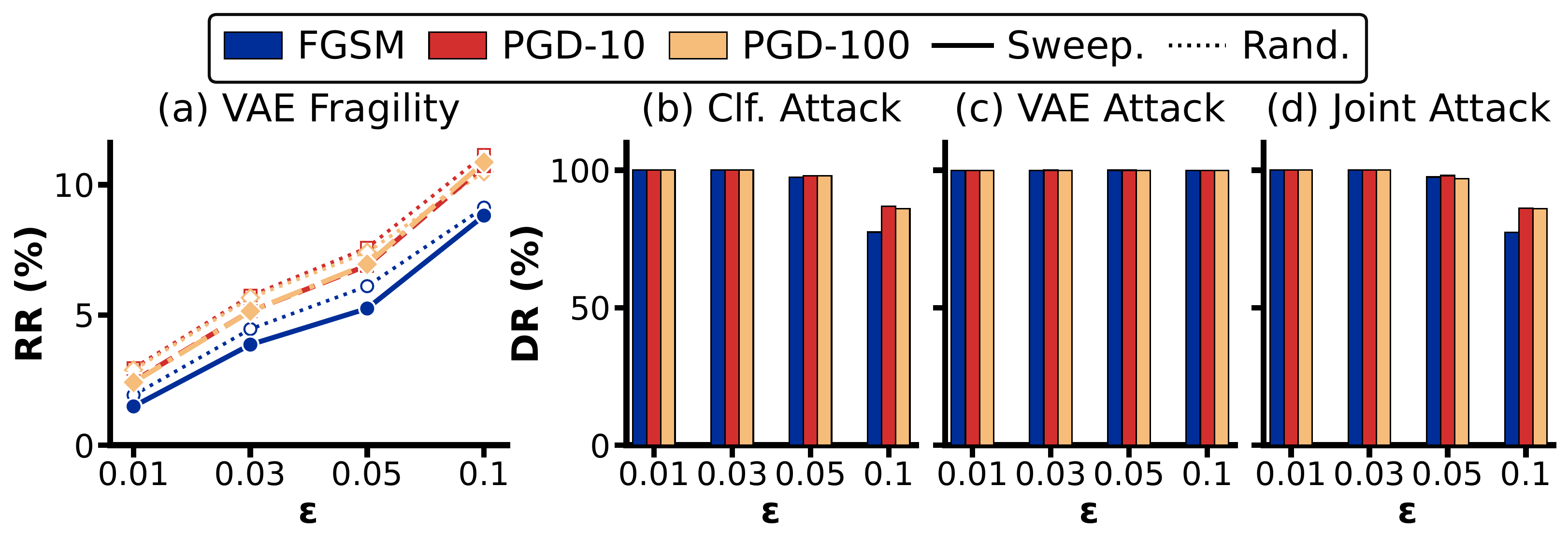}
  \caption{Adversarial dilemma: cross-component safety nets under targeted gradient attacks. (a)~VAE RR under direct recon-loss attack on sweeping and random brute-force zero-day traffic. (b--d)~OOD detection of residual misclassified windows after (b)~classifier-only, (c)~VAE-only, and (d)~joint classifier+VAE attacks on sweeping.}

  \label{fig:wb_dilemma}
\end{figure}

\noindent \textbf{\textit{Cross-component safety nets.}}~The classifier and OOD ensemble operate in parallel within Stage~2 (Section~\ref{sec:ensemble}). A sample must evade \textit{both} branches to escape detection. Figure~\ref{fig:wb_dilemma} probes this redundancy by attacking each branch in turn and then both at once. Panel~(a) shows the VAE reconstruction reduction (RR) under direct gradient attack. Since the VAE serves the OOD detection pathway, we test it on two zero-day categories where reconstruction error is the primary detection signal. At $\varepsilon{=}0.10$, PGD-100 reduces reconstruction error by 10.9\% (sweeping bruteforce) and 10.5\% (random bruteforce), not enough to push the anomaly score below the calibrated threshold. Panel~(b) examines the 2.2\% of sweeping windows that the classifier attack \textit{relabels} as benign. The OOD ensemble flags 86.0\% of them at $\varepsilon{=}0.10$ and all of them at $\varepsilon{\leq}0.03$. The perturbation that shifts the classifier boundary simultaneously \textit{distorts} the latent representation, pushing energy and Mahalanobis scores above their detection thresholds.
Panel~(c) reverses the attack direction. When the full budget targets VAE reconstruction, the classifier remains 99.9\% correct across all $\varepsilon$. Every sample is still classified as an attack. The two loss surfaces share little curvature in common, so a perturbation optimized for one objective produces negligible movement on the other. Panel~(d) tests \textit{budget splitting}. A joint targeted-to-benign plus VAE attack ($\alpha{=}0.5$) still sees 86.1\% of its misclassified samples caught by the OOD ensemble at $\varepsilon{=}0.10$. Dividing the perturbation budget \textit{weakens} both objectives without yielding a complementary gain. We measured the cosine similarity between the classifier and VAE gradient directions across 1,000 samples. The mean is 0.007 with standard deviation 0.058, confirming near-orthogonality in the input space.

\begin{figure}[t]
\centering
\includegraphics[width=\columnwidth]{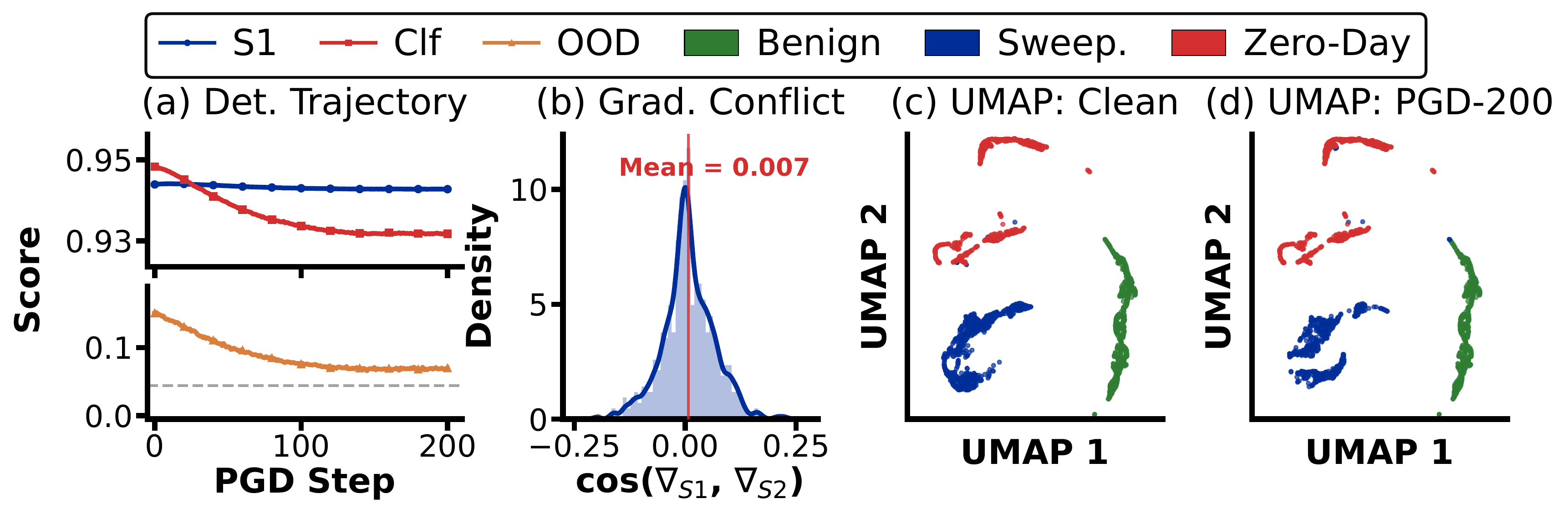}
\caption{Gradient conflict analysis on sweeping at $\varepsilon{=}0.10$. (a)~Detection scores across 200 PGD steps. (b)~Cosine similarity between Stage~1 and Stage~2 gradients. (c)~UMAP of classifier features (no attack). (d)~UMAP under PGD-200 attack.}


\label{fig:wb_analysis}
\end{figure}

\noindent\textbf{\textit{End-to-end synthesis.}}~At $\varepsilon{=}0.10$ on sweeping (the worst case), the joint targeted-to-benign attack produces 2.3\% benign predictions, of which the OOD ensemble catches 74.9\%, leaving an effective pipeline evasion of 0.57\%.
At $\varepsilon{\leq}0.05$, effective evasion is 0.0\% for all categories. Targeting Stage~1 in addition does not \textit{help}. A joint loss $\mathcal{L} = \lambda \mathcal{L}_{S1} + (1{-}\lambda)\mathcal{L}_{S2}$ with $\lambda{=}0.8$ under PGD-200 yields 0.0\% E2E evasion at every $\varepsilon$. Stage~1 and Stage~2 depend on different features 
whose gradient directions \textit{cannot be jointly satisfied} within realizable perturbation bounds.

\noindent \textbf{\textit{Visualization of the gradient conflict.}}~Figure~\ref{fig:wb_analysis} provides four complementary views of why E2E evasion fails. Panel~(a) tracks detection scores across 200 PGD steps on sweeping at $\varepsilon{=}0.10$. The Stage~1 trigger barely moves (0.942$\to$0.941), the classifier confidence drops modestly (0.948$\to$0.927), and the OOD ensemble score plateaus above the anomaly threshold by step~60. Panel~(b) shows the distribution of $\cos(\nabla_x \mathcal{L}_{S1}, \nabla_x \mathcal{L}_{S2})$ across 500 sweeping samples. The mean is 0.007 with standard deviation 0.058, concentrated tightly around zero. The two stages request nearly \textit{orthogonal perturbation directions}, so no single $\delta$ can serve both. Panels~(c--d) visualize the classifier's penultimate feature space via UMAP (Uniform Manifold Approximation and Projection~\cite{mcinnes2018umap}) before and after PGD-200 at $\varepsilon{=}0.10$. The sweeping cluster shifts slightly toward the benign region in panel~(d) compared to the clean embedding in panel~(c), but the three classes remain well \textit{separated}.

\begin{figure}[t]
  \centering
  \includegraphics[width=\columnwidth]{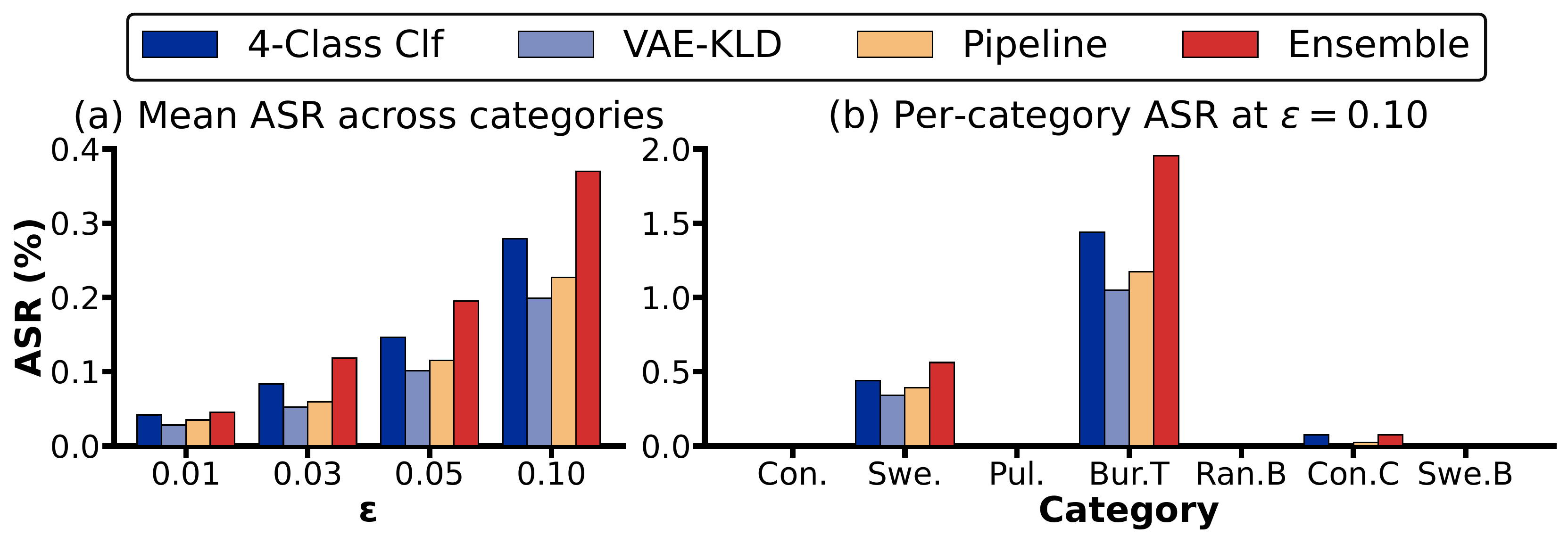}
    \caption{Transfer attack ASR (4,096 samples per category). (a)~Mean ASR across seven zero-day categories vs.\ perturbation budget~$\varepsilon$; (b)~per-category ASR at $\varepsilon{=}0.10$. ASR remains at 0\% for 4 of 7 categories and peaks at 1.95\% on burst triangle.} 
  \label{fig:transfer_asr}
\end{figure}

\subsubsection{Black-Box Attacks ($\mathcal{T}_B$)}
\label{sec:eval_bb}

Under $\mathcal{T}_B$, the adversary has no access to model weights, architecture, or gradients. We evaluate two attack strategies: \textbf{(1)} transfer-based attacks, which train surrogate models and transfer adversarial examples, and \textbf{(2)} query-based attacks, which probe the deployed pipeline's output to guide optimization. All attacks apply \textit{physically realizable perturbation constraints} to every sample.

\noindent \tikz[baseline=(char.base)]\node[shape=circle,draw,inner sep=0.7pt] (char) {1}; \textbf{\textit{Transfer attacks.}}~We train four surrogate architectures that deliberately differ from \textsc{Citadel}'s internal models: a 4-class CNN classifier (75K parameters, 5$\times$5 kernels), a convolutional VAE (600K parameters, latent\_dim=128), a full pipeline surrogate combining a binary Stage~1 approximator with the classifier and VAE, and an ensemble of three classifiers (including a ResNet-18 with 11M parameters). Each surrogate is trained on data from the same distribution and used to craft PGD-50 adversarial examples under physically realizable constraints. Transferred perturbations undergo \textit{channel degradation} 
before evaluation against the real \textsc{Citadel} pipeline, with 4,096 samples per category.
Figure~\ref{fig:transfer_asr} summarizes the results. Figure~\ref{fig:transfer_asr}a shows the mean E2E evasion across all seven categories for each surrogate strategy. Even at $\varepsilon{=}0.10$, the strongest surrogate (ensemble) achieves only 0.37\% mean ASR. The per-category breakdown in Figure~\ref{fig:transfer_asr}b reveals that four of seven categories (constant, pulse, random bruteforce, sweeping bruteforce) achieve \textbf{0.0\%} transfer ASR at every $\varepsilon$. The only category exceeding 1\% is burst triangle (1.95\% at $\varepsilon{=}0.10$, ensemble surrogate), which is also the hardest scenario in clean (non-adversarial) conditions. It is the only zero-day with sub-100\% Stage~1 trigger rate (97.5\%) and the only one with measurable benign leakage (5.1\%). The transfer ``success'' on burst triangle is consistent with its clean-condition miss rate, indicating that the perturbation does not open a new evasion pathway. When targeting Stage~2 in isolation (bypassing Stage~1), transfer ASR is 0.0\% across all 112 configurations, confirming that the multi-signal OOD ensemble is \textit{opaque} to surrogate-based transfer.

\begin{figure}[t]
  \centering
  \includegraphics[width=\columnwidth]{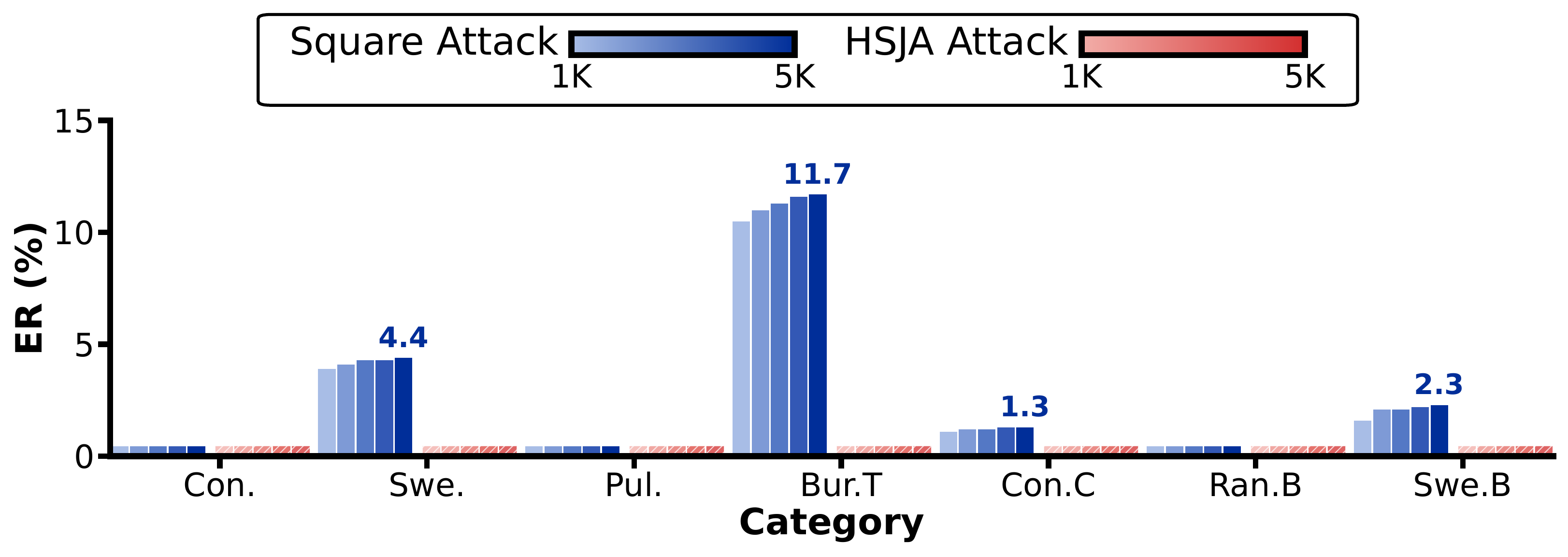}
  \caption{Query-based attack evasion at $\varepsilon{=}0.10$. Blue bars: Square Attack across five query budgets (1K--5K, light to dark). Red bars: HopSkipJump, 0.0\% at all budgets. Stage~2 evasion stays below 0.4\%.}
  
  \label{fig:query_attacks}
\end{figure}

\noindent \tikz[baseline=(char.base)]\node[shape=circle,draw,inner sep=0.7pt] (char) {2}; \textbf{\textit{Query-based attacks.}}~Figure~\ref{fig:query_attacks} compares score-based (Square Attack~\cite{andriushchenko2020square}) and decision-based (HopSkipJump~\cite{chen2020hopskipjump}) attacks at $\varepsilon{=}0.10$ with 4,096 samples per category and query budgets from 1,000 to 5,000 per sample (lower $\varepsilon$ values yield strictly lower evasion). Square Attack reaches 11.0\% on burst triangle and 4.1\% on sweeping at 2,000 queries, with \textit{marginal improvement} up to 5,000 queries (11.7\% and 4.4\%). Three of seven categories remain at 0.0\% regardless of budget. Notably, Square Attack achieves higher per-category evasion than the white-box gradient attacks in Section~\ref{sec:eval_wb}. This is a known phenomenon in multi-component defenses~\cite{tramer2020adaptive}. Gradient-based attackers \textit{must solve a joint optimization} over all components simultaneously, and the near-orthogonal gradient directions between Stage~1 and Stage~2 (Section~\ref{sec:eval_wb}) prevent the optimizer from concentrating its budget on any single component. Score-based query attacks face no such coupling. Square Attack treats the pipeline as a scalar oracle and can allocate perturbation budget \textit{freely}, occasionally finding perturbations that reduce the Stage~1 trigger score without the conflicting Stage~2 gradient pulling in the opposite direction. Stage~2 evasion stays below 0.4\% across all categories and budgets for both attacks, meaning the pipeline evasion comes from Stage~1 boundary cases. HopSkipJump achieves 0.0\% across all 70 configurations. 
As a minimum-distance boundary attack, HopSkipJump searches for the \textit{nearest decision-boundary crossing}. Stage~1's large confidence margin 
places the boundary far from the clean attack distribution, so the nearest adversarial examples require perturbation magnitudes that exceed every $\varepsilon$ we evaluate.

\begin{figure}[t]
  \centering
  \includegraphics[width=\columnwidth]{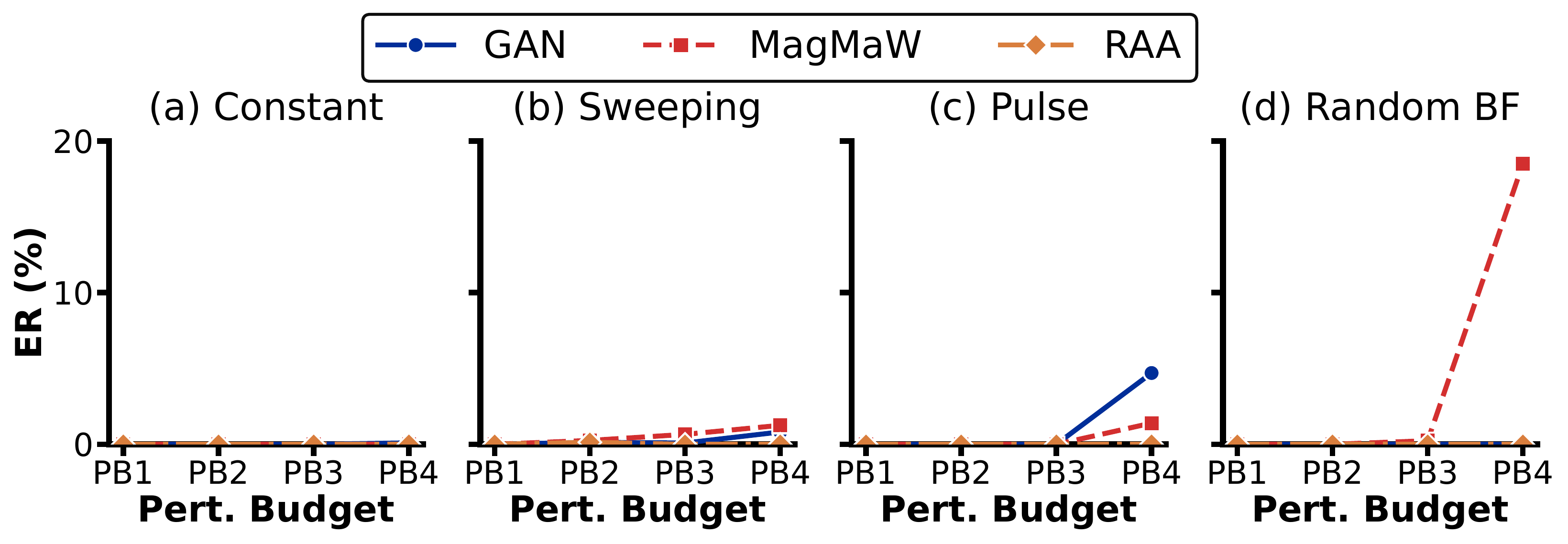}
  \caption{Stage~2 evasion under three SOTA black-box generators.
  Perturbation budgets PB1--PB4 correspond to $\varepsilon \in \{0.01, 0.03, 0.05, 0.10\}$ for GAN and Magmaw, and PSR $\in \{-10, -13, -16, -20\}$\,dB for RAA. Stage~2 evasion remains below 5\% on all known attacks (a--c). Only Magmaw achieves non-trivial evasion on the zero-day category (d), reaching 18.5\% at PB4.}

  \label{fig:sota_s2}
\end{figure}

\subsubsection{State-of-the-Art (SOTA) Adaptive Attacks}
\label{sec:eval_sota}

We evaluate three learned perturbation generators in black-box mode with physically realizable constraints. A GAN-based generator~\cite{alhajjar2021adversarial} that trains an adversarial network to produce \textit{evasion} perturbations, Magmaw~\cite{magmaw2025} which uses manifold-aware projections to \textit{craft} transferable CSI perturbations, and RAA~\cite{bahramali2021robust} which applies \textit{random additive} perturbations with power-level control. Each generator is trained against surrogate models on data from the same distribution and evaluated against \textsc{Citadel}'s full pipeline. Following the pattern established in Section~\ref{sec:eval_bb}, where query-based evasion was traced to Stage~1 boundary cases while Stage~2 remained below 0.4\%, we evaluate Stage~2 resilience as the security-relevant metric. Stage~1 functions as a computational filter (Section~\ref{sec:eval_ablation}). The multi-signal OOD ensemble in Stage~2 is the \textit{security boundary}.
Figure~\ref{fig:sota_s2} shows Stage~2 evasion across three known-attack categories and one zero-day category (random bruteforce) for all three generators. On known attacks (panels~a--c), Stage~2 evasion stays below 4.7\% at every $\varepsilon$, including the extreme budget $\varepsilon{=}0.10$. Constant jamming (panel~a) yields 0.0\% for all three generators across all budgets. The zero-day category random bruteforce (panel~d) separates the generators. Magmaw is the only attack that achieves non-trivial Stage~2 evasion (18.5\% at $\varepsilon{=}0.10$), while GAN and RAA remain below 0.5\%. 
Even under the \textit{strongest published CSI attack (Magmaw)}, Stage~2 evasion on known attacks remains below 4.7\%, confirming that the multi-signal ensemble resists all three attack families. The elevated evasion on random bruteforce (18.5\% at $\varepsilon{=}0.10$) occurs only at an extreme perturbation budget and on a single zero-day category, validating Magmaw as a conservative stress test for the comparative evaluation (Section~\ref{sec:eval_comparative}).


\begin{table}[t]
\centering
\caption{Comparative evaluation against eight baselines on detection performance and adversarial robustness. Magmaw: E2E evasion at $\varepsilon{=}0.05$. Resource costs measured on Jetson Orin Nano (batch~1, FP32).}

\label{tab:comparative}
\resizebox{\columnwidth}{!}{%
\scriptsize
\setlength{\tabcolsep}{3pt}
\begin{tabular}{l r r r r r r r}
\toprule
& \multicolumn{4}{c}{\textbf{Detection Performance}} & \multicolumn{3}{c}{\textbf{Resource Cost}} \\
\cmidrule(lr){2-5} \cmidrule(l){6-8}
\textbf{Method}
  & \textbf{Known}
  & \textbf{ZD}
  & \textbf{FPR}
  & \textbf{Magmaw}
  & \textbf{Params}
  & \textbf{Lat.}
  & \textbf{Energy} \\
  & \textbf{DR} & \textbf{DR} & & \textbf{(\%$\downarrow$)}
  & & \textbf{(ms)} & \textbf{(mJ)} \\
\midrule
\textsc{Citadel} (S1)
  & -- & -- & -- & --
  & 1.4K & \textbf{1.1} & \textbf{6.5} \\
\textsc{Citadel} (E2E)
  & \textbf{100.0} & \textbf{97.1} & \textbf{0.4} & \textbf{4.2\,(S2:0.2)}
  & 2,845K & 14.2 & 95.9 \\
\midrule
\multicolumn{8}{l}{\textit{Post-hoc OOD detectors}} \\
\addlinespace[2pt]
MSP~\cite{hendrycks2017}
  & 32.7 & 38.8 & 20.0 & 56.1
  & 423K\rlap{$^\dagger$} & 2.0 & 12.6 \\
KNN-OOD~\cite{sun2022}
  & 44.8 & 50.2 & 20.8 & 41.7
  & 423K\rlap{$^\dagger$} & 2.0 & 12.5 \\
ASH-S~\cite{djurisic2023}
  & 46.7 & 44.0 & 24.3 & 42.3
  & 423K\rlap{$^\dagger$} & 2.0 & 12.5 \\
ViM~\cite{wang2022vim}
  & 4.7  & 1.9  & 9.7  & 96.6
  & 423K\rlap{$^\dagger$} & 2.0 & 12.5 \\
\midrule
\multicolumn{8}{l}{\textit{Domain-specific detectors}} \\
\addlinespace[2pt]
JADE~\cite{kilinc2022}
  & 64.0 & 82.8 & 13.2 & 23.5
  & 154K & 1.3 & 7.3 \\
BloodHound+~\cite{sciancalepore2023}
  & 35.9 & 26.4 & 12.4 & 67.6
  & 47K & 1.2 & 6.8 \\
JamShield~\cite{panitsas2025}
  & 99.1 & 44.8 & \textbf{0.1} & 92.3
  & 423K\rlap{$^\dagger$} & 2.0 & 12.6 \\
HussainEdge~\cite{hussain2022}
  & 99.9 & 55.9 & 5.8  & 48.4
  & 913K & 0.9 & \textbf{5.3} \\
\bottomrule
\multicolumn{8}{l}{$^\dagger$Shared CSI2DClassifier backbone (422.5K params); post-hoc scoring adds negligible overhead.} \\
\end{tabular}
}%
\end{table}


A key architectural property reinforces these results.
Stage~1 executes \textit{on the sensor node itself},
processing CSI internally before any data leaves the
device. An RF-domain attacker has no channel to
selectively suppress Stage~1 without simultaneously
altering the CSI measurements that Stage~2 analyzes.
The generators that achieve high E2E bypass rates
(e.g., Magmaw at 57.9\% on sweeping at
$\varepsilon{=}0.10$) do so by suppressing the
broadband spectral energy that Stage~1 monitors,
causing suspicious windows to \textit{not be escalated}.
However, this is a computational shortcut, not a
security breach. Any attack strong enough to
compromise the protected link must perturb CSI, and
once escalated, Stage~2 detects it with less than
0.2\% evasion. Stage~1 bypass without Stage~2 evasion
means the attack either goes undetected because it is
too weak to affect the link, or is caught the moment
it becomes operationally relevant. The security
boundary is Stage~2, and no generator breaches it
within realizable budgets.

\subsection{Comparative Evaluation}
\label{sec:eval_comparative}


To contextualize \textsc{Citadel}'s performance, we compare against eight detection methods (Table~\ref{tab:comparative}) spanning four categories: \textit{post-hoc} OOD detectors originally designed for image classification (MSP~\cite{hendrycks2017}, KNN-OOD~\cite{sun2022}, ASH-S~\cite{djurisic2023}, ViM~\cite{wang2022vim}), \textit{reconstruction-based} anomaly detectors (JADE~\cite{kilinc2022}, BloodHound+~\cite{sciancalepore2023}), and \textit{signal-level} WiFi detectors (JamShield~\cite{panitsas2025}, HussainEdge~\cite{hussain2022}). All methods are re-implemented on CSI data using the same training set and evaluation protocol. Post-hoc methods use two-class backbone on normal traffic, domain-specific methods use four-class supervised backbone. 



\textsc{Citadel} is the only method that achieves high detection across all three dimensions. HussainEdge (99.9\%) and JamShield (99.1\%) approach \textsc{Citadel}'s 100\% on known attacks but collapse on zero-day scenarios (55.9\% and 44.8\% respectively). Conversely, JADE achieves the highest baseline zero-day rate (82.8\%) but detects only 64.0\% of known attacks and suffers 23.5\% evasion under Magmaw. The post-hoc OOD methods \textit{fail} to transfer to the CSI domain. ViM achieves only 1.9\% zero-day detection and 96.6\% Magmaw evasion, confirming that virtual-logit augmentation \textit{does not generalize} to spectrotemporal data.

\begin{figure}[t]
    \centering
    \includegraphics[width=\columnwidth]{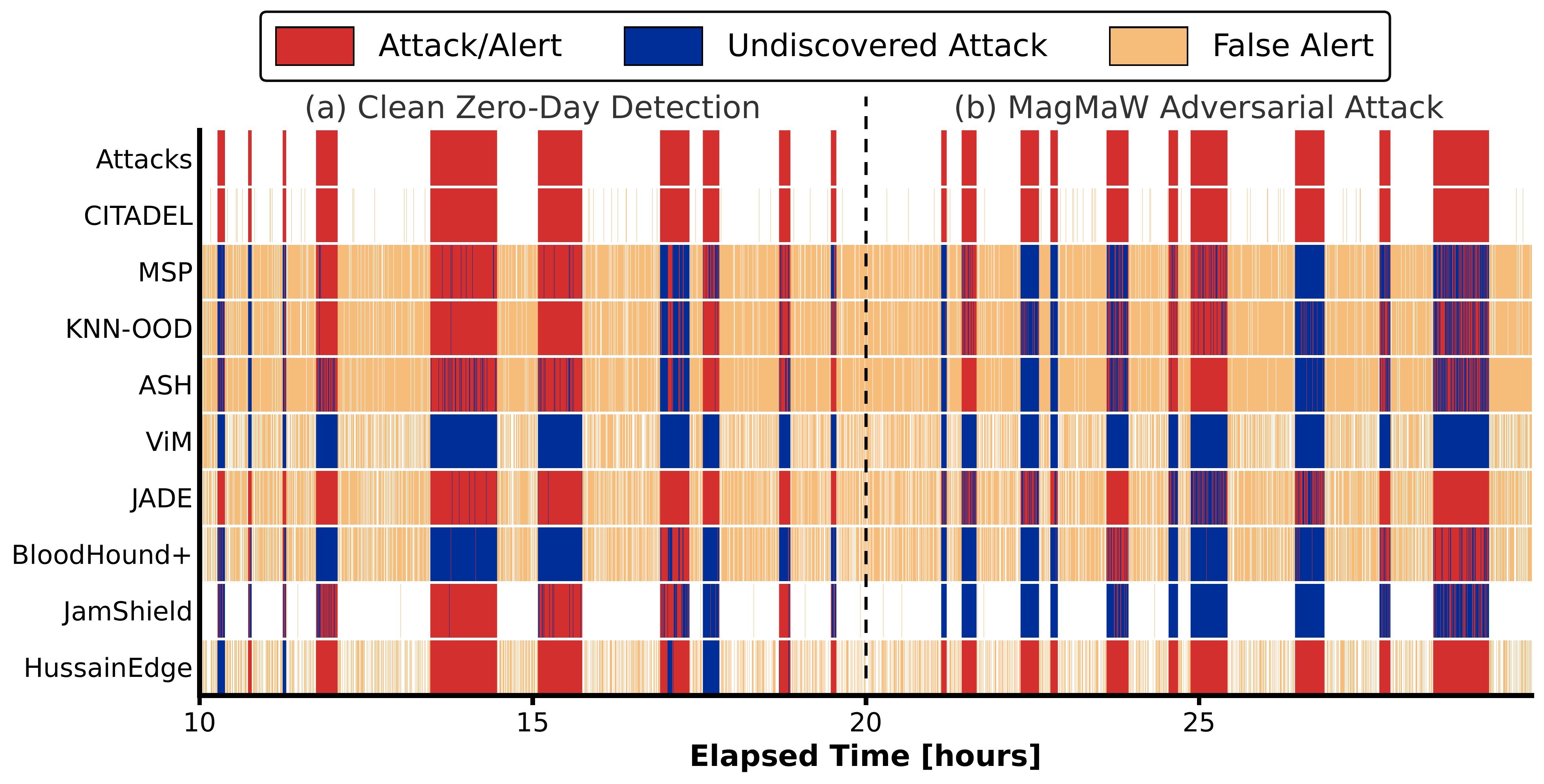}
     
    
    \caption{15-hour operational timeline from testbed data. Each row represents one detection method. (a)~Zero-day attacks without adversarial perturbation; \textsc{Citadel} detects all injected attacks while baselines miss up to 100\% of zero-day scenarios. (b)~Known attacks under Magmaw ($\varepsilon{=}0.05$); \textsc{Citadel} maintains 4.2\% E2E evasion. 
    }
  
  \label{fig:alert_timeline}
\end{figure}

Under Magmaw at $\varepsilon{=}0.05$, \textsc{Citadel}'s E2E evasion (4.2\%) is \textit{6 times lower} than JADE (23.5\%) and \textit{22 times lower} than JamShield (92.3\%). The 4.2\% average includes sweeping (16.3\%), where the generator learns to suppress Stage~1's trigger, while constant, pulse, and random bruteforce remain below 0.5\%. At the Stage~2 security boundary, evasion drops to 0.2\% (Section~\ref{sec:eval_sota}). Three failure modes explain this. Confidence-based methods rely on a \textit{single discriminative pathway} that the generator can target \textit{without} constraint. Reconstruction-based methods offer partial robustness but a lone reconstruction signal gives the attacker a \textit{single optimization target}. Signal-level methods (JamShield, HussainEdge) operate on \textit{aggregate statistics} sensitive to small CSI modifications within realizable bounds. \textsc{Citadel}'s multi-signal ensemble forces the attacker to satisfy conflicting objectives \textit{simultaneously}, and no generator achieves this within realizable budgets.

Figures~\ref{fig:alert_timeline} provide an operational perspective, monitoring a 15-hour session. In Figure~\ref{fig:alert_timeline}a, \textsc{Citadel}'s alert stream mirrors the ground truth while post-hoc methods produce near-continuous \textit{false alerts}. Under Magmaw (Figure~\ref{fig:alert_timeline}b), \textsc{Citadel} remains intact while JamShield collapses to near-total evasion. No baseline maintains both detection completeness and false-alarm control under adversarial conditions.

\noindent \textbf{\textit{Resource cost.}}~\textsc{Citadel}'s two-stage design amortizes its cost. Stage~1 runs on every window in 1.1\,ms at 6.5\,mJ on the Jetson, filtering 92\% of benign traffic before Stage~2 is \textit{invoked}. On the target ESP32-C6 microcontroller, the same model completes inference in 9.4\,ms within 17\,KiB of SRAM (Appendix~\ref{app:esp32c6}). Worst-case E2E cost is 14.2\,ms at 95.9\,mJ, with 20.4\,MB peak memory and 0.167\,GFLOPs. Single-stage baselines are faster but at substantially \textit{weaker} detection and robustness. Stage~2 overhead is dominated by two classifier passes (4.1\,ms), VAE encode/decode (3.9\,ms), and diffusion denoising (4.4\,ms), completing well within the 250\,ms inter-window budget at 4\,Hz CSI sampling (\textbf{RQ4}).

\section{Limitations and Discussion}
\label{sec:discussion}

\noindent\textbf{Generalization Beyond Controlled Environments.}~Our evaluation was conducted in a controlled laboratory with a single jammer model (HackRF One) and fixed sensor topology. Production IIoT floors introduce metallic reflectors, co-channel interference, and environmental non-stationarities (e.g., shift changes, moving personnel) that the current training distribution does \textit{not fully capture}.
While the system's reliance on the physical invariant (jamming inevitably perturbs CSI) is \textit{environment-agnostic}, the calibrated OOD thresholds are environment-specific. Deploying \textsc{Citadel} in a new facility would require a site-specific calibration phase, analogous to the commissioning step common in industrial monitoring systems. Future work should validate detection performance through \textit{field trials} in operational factories and investigate online recalibration strategies that adapt to environmental drift without retraining.

\noindent\textbf{Spectral and Temporal Coverage.}~\textsc{Citadel} monitors a single 20\,MHz channel at approximately 4\,Hz, imposing two coverage boundaries. Spectrally, detection is confined to the monitored channel. Facilities operating multiple access points on non-overlapping channels would require per-channel deployment to achieve \textit{full spectral coverage}. Correlating CSI streams across access points would introduce spatial and spectral diversity to \textit{broaden} protection. Temporally, the 8-second observation window means sub-100\,ms micro-bursts may produce insufficient distortion to \textit{trigger} the ensemble. Higher acquisition rates on Wi-Fi~6/6E chipsets or 5G New Radio reference signals~\cite{lichtman2016lte} would improve temporal resolution and broaden applicability beyond the current testbed.

\noindent\textbf{Simulated vs.\ Over-the-Air Adversarial Validation.}~The adversarial evaluation enforces \textit{physical realizability} through differentiable projections (C1--C3) in the digital domain. These constraints capture necessary conditions for over-the-air feasibility. A real-time adversarial transmitter would face additional \textit{impairments} that further degrade the attacker's effective budget. The sub-2\% gradient-based evasion reported in Section~\ref{sec:eval_wb} is therefore likely a conservative estimate of deployed robustness, but validating this requires closed-loop experiments with a second SDR transmitting adversarial waveforms while \textsc{Citadel} operates in real time.
The construction of such a testbed, along with a standardized benchmark for physically realized adversarial attacks on wireless ML systems, represents an important \textit{direction} for the broader community.

\section{Related Work}
\label{sec:related}

\noindent\textbf{Jamming detection.}
Section~\ref{sec:introduction} compared four sensing modalities (Figure~\ref{fig:modalities}). Here we position \textsc{Citadel} against the systems built on them. RSS-based~\cite{hussain2022} and cross-layer~\cite{panitsas2025} detectors achieve high known-attack accuracy but offer \textit{no path} to zero-day generalization. Reconstruction-based approaches~\cite{kilinc2022, sciancalepore2023} provide zero-day capability through deviation scoring but \textit{sacrifice} discriminative precision on known attacks. At the physical layer, I/Q spectrogram classifiers~\cite{varotto2024, hanegraaf2025} demonstrated supervised detection \textit{without} addressing unseen waveforms or adversarial conditions. The first CSI-based study on ESP32~\cite{11096274} confirmed jamming sensitivity without building a complete pipeline.  SpotLight~\cite{sun2024spotlight} proposed generative anomaly detection for Open RAN but targets cellular infrastructure rather than IIoT. \textit{No prior system} unifies known-attack classification, zero-day OOD detection, adversarial evaluation, and hardware deployment.

\noindent\textbf{Adversarial attacks on wireless systems.}
The adversarial challenge (\textbf{RQ3}) motivated physically constrained evaluation. We detail the constraint \textit{gap} in existing attack models. RAA~\cite{bahramali2021robust} bounds perturbation power but does not enforce subcarrier correlation or temporal coherence. Magmaw~\cite{magmaw2025} adds OFDM-aware preamble corruption but \textit{omits} indoor-propagation constraints. Practical over-the-air attacks~\cite{li2024practical, liu2023rafa, zhou2023ristealth} demonstrate feasibility but target single-model defenses. Phantom-CSI~\cite{he2023phantom} further showed that adversarially crafted CSI patterns can \textit{fool} liveness detectors. All these works highlight that unconstrained digital perturbations ignore physical propagation \textit{effects}, motivating the constrained evaluation methodology~\cite{chernikova2022fence, carlini2019evaluating, tramer2020adaptive} that \textsc{Citadel} adopts.

\noindent\textbf{OOD detection and adversarial-resilient systems.}
The OOD methods evaluated in Section~\ref{sec:eval_comparative} were designed for high-dimensional image data. Our results confirm that they \textit{do not transfer} to the temporally correlated structure of CSI. Diffusion models have been applied to adversarial purification~\cite{nie2022diffpure} and anomaly detection~\cite{ho2020ddpm}, but only in the image domain. In network intrusion detection, MANDA~\cite{wang2022manda} fuses manifold inconsistency with boundary scoring but \textit{requires} adversarial training data, NIDS-DA~\cite{kumar2025nidsda} operates at high FPR, and AdvPurRec~\cite{alhussien2024advpurrec} achieves limited recovery on tabular traffic. All operate on network-flow features. None are evaluated under wireless-specific attacks. The closest architectural parallel is the multi-stage IDS of~\cite{verkerken2023hierarchical}. \textsc{Citadel} extends this paradigm to the physical layer with a microcontroller-deployable first stage, a three-signal OOD ensemble calibrated without zero-day data, and evaluation under physically realizable constraints.

\section{Conclusion}
\label{sec:conclusion}

We presented \textsc{Citadel}, a two-stage hierarchical system for detecting RF jamming attacks against \textit{wireless links} in IIoT environments using only CSI (\textbf{RQ1}). By decomposing detection across hardware tiers, a \textit{lightweight} binary trigger on ESP32 microcontrollers at the field level and a multi-signal OOD \textit{ensemble} on an edge GPU at the supervisory level, \textsc{Citadel} achieves both the computational efficiency required for real-time edge deployment and the detection depth needed to identify previously \textit{unseen} attack strategies. On 15 zero-day scenarios (\textbf{RQ2}), the system detects 97.1\% of attack windows as anomalous while maintaining a 0.4\% E2E FPR, demonstrating that the fusion of KL divergence, energy scoring, and Mahalanobis distance in orthogonal information spaces provides robust generalization without any zero-day supervision. Under adversarial (\textbf{RQ3}) evaluation spanning white-box and black-box threat models, gradient-based evasion stays below 2\% at every tested perturbation budget, while the strongest published CSI attack generator achieves less than 5\% average evasion, confirming that the two-stage architecture imposes \textit{conflicting} optimization objectives that prevent simultaneous evasion of both pipeline stages. The entire pipeline runs in real time on edge hardware (\textbf{RQ4}), completing end-to-end inference in 14.2\,ms at 95.9\,mJ.

\bibliographystyle{IEEEtran}
\bibliography{references}

\appendix

\section{Ethical Considerations}
\label{app:ethics}

This work involves the deliberate generation of RF jamming signals, which raises ethical considerations regarding spectrum interference and potential misuse.

\noindent \textbf{Controlled environment.}~All RF jamming experiments were conducted in a controlled laboratory environment at low to moderate power levels (IF gain 10--20\,dB, well below the HackRF One's 47\,dB hardware maximum, with the RF amplifier permanently disabled). No production IIoT systems, public WiFi networks, or third-party communications were disrupted at any point during data collection.

\noindent \textbf{Defensive purpose.}~The jamming attack tooling and adversarial evaluation framework are developed and released solely to enable reproducible evaluation of defensive detection systems. The attack scenarios are designed to stress-test \textsc{Citadel} and baseline detectors under realistic conditions, following established responsible-disclosure practices in the wireless security community.

\noindent \textbf{Dual-use considerations.}~While the jamming scripts and adversarial attack implementations could theoretically be repurposed for offensive use, the techniques they implement (SDR-based interference, gradient-based evasion) are already well-documented in the public literature~\cite{ali2022jamrf, bahramali2021robust, magmaw2025}. Releasing our implementations alongside the defensive system provides net benefit by enabling the community to evaluate and improve jamming defenses.

\section{Man-in-the-Middle Threat Analysis}
\label{app:mitm}

In addition to the threat models evaluated in the main body, we define a man-in-the-middle threat model ($\mathcal{T}_{\text{MITM}}$) for completeness. The adversary compromises the gateway between Stage~1 and Stage~2, modifying CSI tensors in the digital domain after RF-to-digital conversion. The attacker can inject arbitrary digital perturbations but cannot retroactively alter the Stage~1 decision already made on the sensor node, testing Stage~2's independent robustness in isolation. Our evaluation confirmed that replay attacks achieve 93.7\% evasion while combined attacks reach 100\%, but random perturbations remain limited to 13.5\% maximum evasion, validating Stage~2's resilience against non-replay digital manipulation.

Protocol-level countermeasures mitigate $\mathcal{T}_{\text{MITM}}$ independently of the ML pipeline: mutual TLS~v1.3 authentication between sensor nodes and the edge tier prevents unauthorized gateway substitution; a periodic heartbeat watchdog (configurable interval, default 5\,s) detects link interruption consistent with active interception; and cryptographic nonces attached to each CSI window enable the edge tier to reject replayed or reordered measurements. These defenses operate at the transport layer and are orthogonal to \textsc{Citadel}'s detection pipeline.

\section{Dataset Statistics}
\label{app:dataset}

Table~\ref{tab:app_dataset} provides the complete dataset breakdown.

\begin{table}[H]
\centering
\caption{Dataset composition and split sizes. All splits are temporal with 128-window data-split purge gaps to prevent autocorrelation leakage.}
\label{tab:app_dataset}
\scriptsize
\setlength{\tabcolsep}{3pt}
\adjustbox{max width=\columnwidth}{%
\begin{tabular}{lrl}
\toprule
\textbf{Category} & \textbf{Windows} & \textbf{Notes} \\
\midrule
\multicolumn{3}{l}{\textit{Benign Traffic (264,014)}} \\
\quad Quiet & 203,536 & In-distribution \\
\quad Movement & 60,478 & OOD by design \\
\midrule
\multicolumn{3}{l}{\textit{Known Attacks (18 scenarios, 366,796)}} \\
\quad Constant (6) & ${\sim}$122k & 2 waveforms $\times$ 3 powers \\
\quad Sweeping (6) & ${\sim}$122k & 2 waveforms $\times$ 3 powers \\
\quad Pulse (6) & ${\sim}$122k & 2 waveforms $\times$ 3 powers \\
\midrule
\multicolumn{3}{l}{\textit{Zero-Day (15 scenarios, 184,508)}} \\
\quad Novel timing (8) & Varies & Random + Burst \\
\quad Novel waveform (7) & Varies & Known timing, new wfm \\
\midrule
\textbf{Total} & $>$815k & \\
\addlinespace
\multicolumn{3}{l}{\textit{Data Split}} \\
\quad Train & 70\% & Temporal split \\
\quad Validation & 15\% & 5-fold blocked CV \\
\quad Test & 15\% & Final evaluation \\
\bottomrule
\end{tabular}}
\end{table}

\section{Zero-Day Coverage Matrix}
\label{app:coverage}

\begin{table}[H]
\centering
\caption{Zero-day coverage matrix. K\,=\,known (training), Z\,=\,zero-day (test). The 15 zero-day scenarios span two novelty axes: novel timing (random/burst columns) and novel waveforms on known timing.}
\label{tab:coverage}
\scriptsize
\setlength{\tabcolsep}{3pt}
\adjustbox{max width=\columnwidth}{%
\begin{tabular}{lcccccc}
\toprule
\textbf{Waveform} & \textbf{Const.} & \textbf{Sweep} & \textbf{Pulse} & \textbf{Rand.} & \textbf{Burst} & \textbf{Total} \\
\midrule
Gaussian   & K & K & K & -- & -- & 3K \\
QPSK       & K & K & K & -- & -- & 3K \\
Bruteforce & -- & Z & Z & Z & -- & 3Z \\
Chirp      & Z & -- & Z & Z & Z & 4Z \\
FSK        & -- & Z & -- & Z & -- & 2Z \\
Sawtooth   & -- & Z & -- & Z & -- & 2Z \\
Square     & -- & -- & -- & Z & -- & 1Z \\
Triangle   & -- & -- & Z & Z & Z & 3Z \\
\midrule
\textbf{Total} & \textbf{6K+1Z} & \textbf{6K+3Z} & \textbf{6K+3Z} & \textbf{6Z} & \textbf{2Z} & \textbf{18K+15Z} \\
\bottomrule
\end{tabular}}
\end{table}

\section{Training Convergence}
\label{app:training}

\begin{figure}[H]
\centering
\includegraphics[width=\columnwidth]{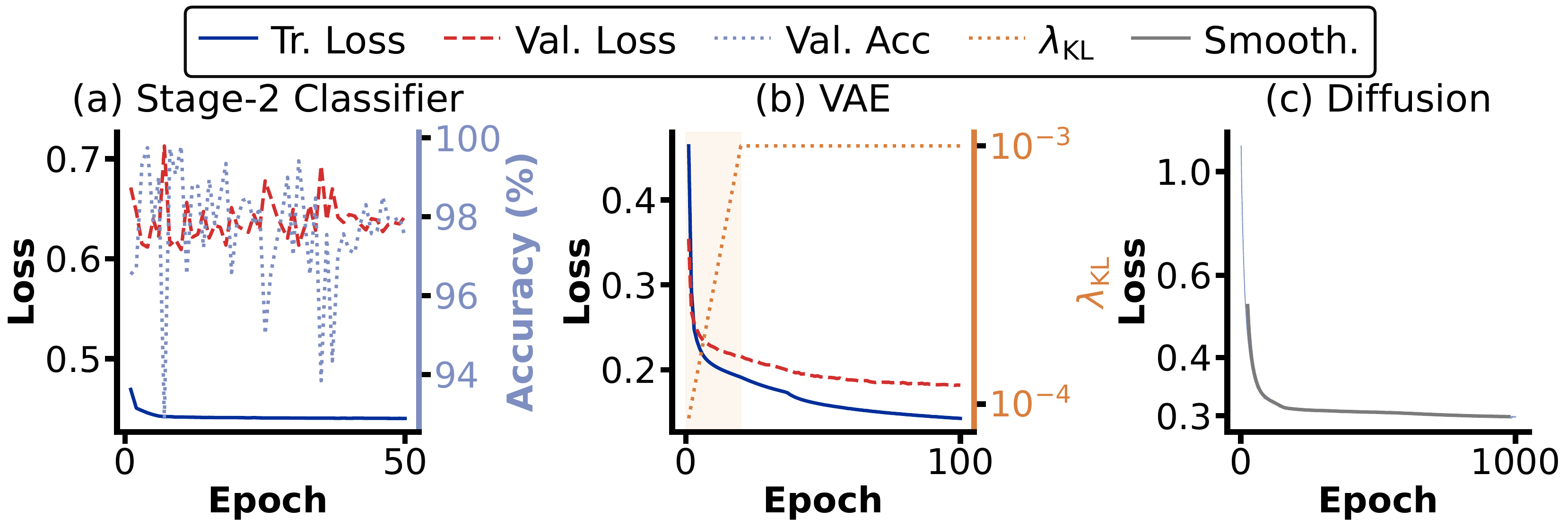}
\caption{Training convergence. (a)~Classifier loss and accuracy over 50 epochs. (b)~VAE loss with KL weight warmup over 100 epochs. (c)~Diffusion DDPM loss over 1{,}000 epochs (log scale).}
\label{fig:training_curves}
\end{figure}

Figure~\ref{fig:training_curves} shows the training dynamics of the three Stage~2 components.

\section{Calibration Configuration}
\label{app:calibration}

Table~\ref{tab:app_calibration} lists the OOD detection calibration parameters determined via 5-fold blocked temporal cross-validation.

\begin{table}[H]
\centering
\caption{OOD calibration parameters. AllF thresholds from out-of-fold (OOF) scores via 5-fold blocked temporal CV with 256-window K-fold purge gaps (larger than data-split gaps for stricter fold isolation).}
\label{tab:app_calibration}
\scriptsize
\adjustbox{max width=\columnwidth}{%
\begin{tabular}{lll}
\toprule
\textbf{Parameter} & \textbf{Method} & \textbf{Value} \\
\midrule
S1 threshold ($\tau_1$) & P90 normal val. & 0.293 \\
Quiet anomaly thresh. & P90 OOF quiet & 0.7456 \\
Attack anomaly thresh. & P95 OOF attack & 0.2698 \\
KLD extreme override & P99.5 OOF & 3.1873 \\
Ensemble weights & Equal & $\frac{1}{3}$, $\frac{1}{3}$, $\frac{1}{3}$ \\
Diffusion timestep ($t$) & Sweep (28 configs) & 10 \\
Temperature ($T$) & Sweep (28 configs) & 0.5 \\
KLD soft scoring & P98 upper bound & Decoupled \\
\bottomrule
\end{tabular}}
\end{table}

\section{OOD Signal Independence}
\label{app:signal_correlation}

\begin{figure}[H]
\centering
\includegraphics[width=\columnwidth]{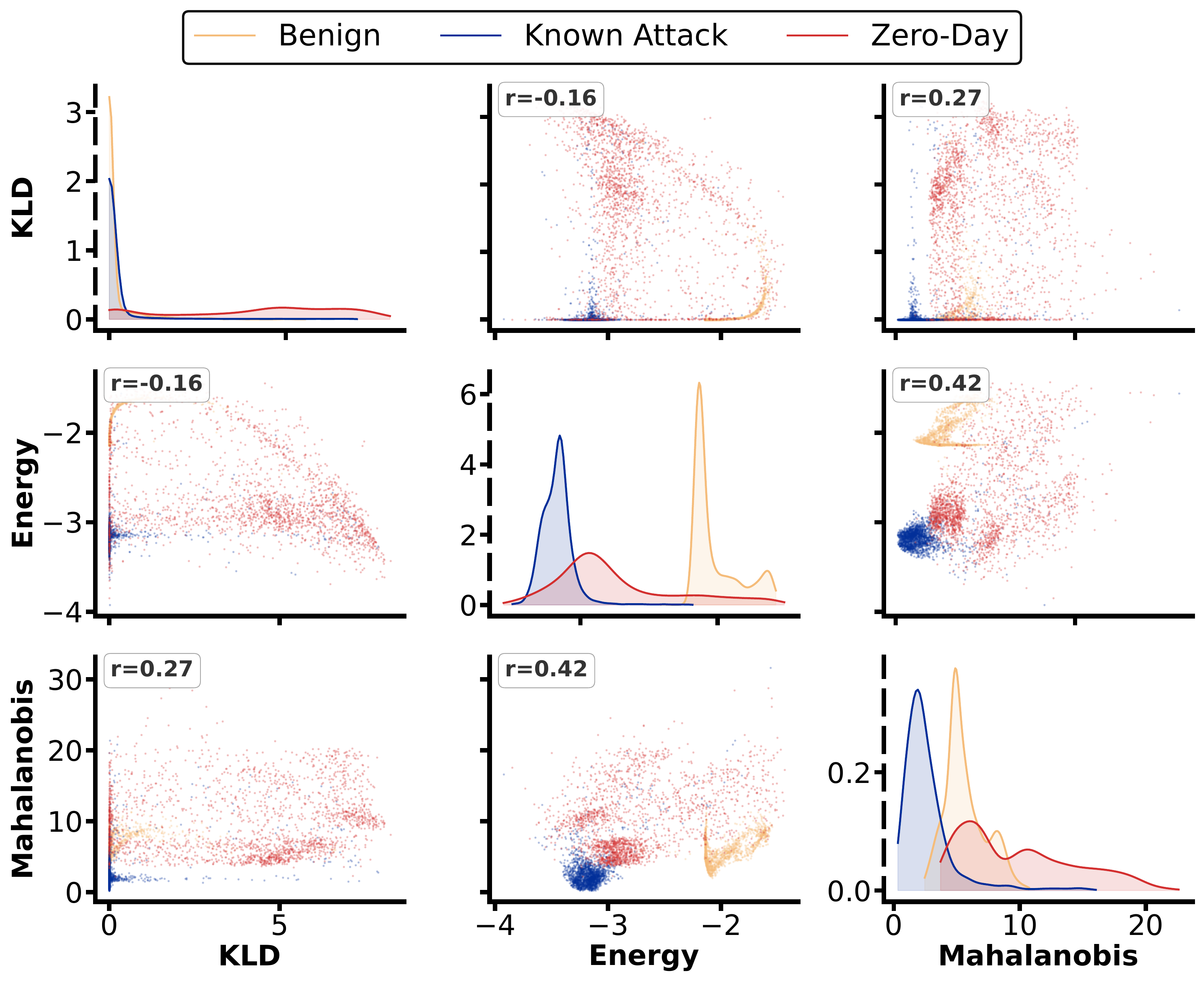}
\caption{Pairwise correlation of the three OOD signals across all traffic classes (triggered samples only). Diagonal panels show per-class KDE distributions; their dashed y-axes denote probability density (unitless). Off-diagonal panels show pairwise scatter plots with Pearson $r$ annotated. Low inter-signal correlations confirm that KLD, Energy, and Mahalanobis capture complementary information.}
\label{fig:signal_correlation}
\end{figure}

Figure~\ref{fig:signal_correlation} shows the pairwise scatter matrix of the three OOD signals (triggered samples only). Diagonal panels display per-class KDE distributions. Off-diagonal panels show pairwise scatter plots with Pearson correlation coefficients annotated. The correlations are consistently low across all pairs:

\begin{itemize}[leftmargin=1.2em,itemsep=2pt]
  \item \textbf{KLD vs.\ Energy} ($r = -0.16$): weakly negative.
    High KLD (VAE reconstruction failure) does not imply low classifier
    confidence; these signals operate in different representation
    spaces (latent vs.\ logit).
  \item \textbf{KLD vs.\ Mahalanobis} ($r = 0.27$): weakly positive.
    Both increase for zero-day attacks but through different mechanisms
    (reconstruction divergence vs.\ feature-space distance).  For
    zero-day traffic specifically, this correlation drops to $r = 0.04$,
    confirming near-independence on the target class.
  \item \textbf{Energy vs.\ Mahalanobis} ($r = 0.42$): moderate.
    Both rely on the classifier's feature space, explaining the
    highest pairwise correlation.  Even so, $r^2 = 0.18$, meaning
    82\% of variance is unshared.
\end{itemize}

These low correlations justify the equal-weight ($\frac{1}{3}$, $\frac{1}{3}$, $\frac{1}{3}$) fusion strategy: since the signals are largely non-redundant, a weighted combination captures strictly more information than any single signal.  Optimizing the weights via grid search yielded no significant improvement over equal weighting, consistent with the low redundancy observed here.

\section{Stage~1 On-Device Inference on the ESP32-C6}
\label{app:esp32c6}

The main body reports E2E inference cost on the edge GPU. This appendix provides the complementary sensor-tier measurement. Stage~1 executing on the same ESP32-C6 microcontroller that acts as the CSI sensor node in our testbed (Section~\ref{sec:implementation}), substantiating the microcontroller-feasibility component of \textbf{RQ4}. The Stage~1 TinyClassifier ($1{,}362$ parameters) is compiled to a flash-resident \texttt{int8} model via batch-normalization folding into the convolutional weights followed by full \texttt{int8} post-training quantization with a $400$-window stratified representative set, and executed on the ESP32-C6 (RV32IMAC single core at $160$\,MHz, ${\sim}$$416$\,KiB of SRAM) under two independent microcontroller runtimes: TFLM 1.3
and ESP-DL v3.3.0 with scalar reference kernels for the RV32IMAC core. Both runtimes reproduce the host-side \texttt{int8} simulator bit-exactly on the embedded sanity set.

\noindent \textbf{Numerical fidelity.}~Table~\ref{tab:app_esp32c6_constraints} reports Stage~1 test-set performance across three numerical backends, with the threshold $\tau$ retuned on the validation split for each backend under the joint FPR/FNR below 5\% constraint. The TFLite float32 backend is bit-equivalent to the folded PyTorch reference ($\max |\Delta p| = 7{\times}10^{-7}$). Per-sample probabilities under \texttt{int8} quantization shift by at most $|\Delta p| {\approx} 0.40$ on ambiguous samples yet preserve the argmax across the test split, and the dual FPR/FNR constraint is satisfied with more than an order of magnitude of margin on the ESP32-C6.

\noindent \textbf{On-device inference performance.}~Latency on the ESP32-C6 is measured with the hardware timer \texttt{esp\_timer\_get\_time()} ($1$\,\textmu s resolution) over $1{,}000$ timed inferences preceded by $20$ warmup iterations. Table~\ref{tab:app_esp32c6_runtime} summarizes the median and tail latency, on-chip memory occupancy, and throughput for both runtimes. Stage~1 fits in under $17$\,KiB of SRAM for either runtime, preserving the memory budget required by the ESP32-C6 Wi-Fi and TCP/IP stack and by the CSI acquisition pipeline co-resident on the sensor node. Median latency is $16.7$\,ms under TFLM and $9.4$\,ms under ESP-DL. Both are well within the $250$\,ms inter-window budget imposed by the $4$\,Hz CSI stream, leaving $15$--$27$ times headroom for the sensor-node workload. The ESP-DL throughput of $105.8$ inferences per second per core is sufficient to support multiple CSI-capable radios polled from a single ESP32-C6.

\begin{table}[H]
\centering
\caption{Stage~1 dual-constraint verification across numerical backends on the test split. $\tau$ is retuned on the validation split under the joint FPR/FNR below 5\% constraint.}
\label{tab:app_esp32c6_constraints}
\scriptsize
\adjustbox{max width=\columnwidth}{%
\begin{tabular}{lrrrr}
\toprule
\textbf{Backend} & \textbf{$\tau$} & \textbf{FPR} & \textbf{FNR} & \textbf{TPR} \\
\midrule
PyTorch folded (f32)   & 0.5770 & 0.366\% & 0.555\% & 99.445\% \\
TFLite (f32)           & 0.5770 & 0.366\% & 0.555\% & 99.445\% \\
TFLite (\texttt{int8}) & 0.5690 & 0.414\% & 0.474\% & 99.526\% \\
\bottomrule
\end{tabular}}
\end{table}

\begin{table}[H]
\centering
\caption{Stage~1 on-device inference on the ESP32-C6 (single RV32IMAC core at $160$\,MHz). Latency is measured over $1{,}000$ iterations after $20$ warmup iterations; throughput is reported per core.}
\label{tab:app_esp32c6_runtime}
\scriptsize
\setlength{\tabcolsep}{5pt}
\adjustbox{max width=\columnwidth}{%
\begin{tabular}{lrr}
\toprule
\textbf{Metric} & \textbf{TFLM} & \textbf{ESP-DL} \\
\midrule
\multicolumn{3}{l}{\textit{Latency}} \\
\quad Median p50 (\textmu s)         & 16{,}703     & \textbf{9{,}437}  \\
\quad Mean (\textmu s)               & 16{,}703.0   & 9{,}458.1 \\
\quad Tail p99 (\textmu s)           & 16{,}780     & 9{,}645   \\
\midrule
\multicolumn{3}{l}{\textit{On-chip memory}} \\
\quad Model flash (bytes)            & 6{,}008      & 6{,}832   \\
\quad SRAM (bytes)                   & 8{,}924      & 16{,}732  \\
\midrule
\multicolumn{3}{l}{\textit{Throughput}} \\
\quad Inferences per second          & 59.9         & \textbf{105.8} \\
\bottomrule
\end{tabular}}
\end{table}

Together with the edge-GPU figures of Section~\ref{sec:implementation}, these measurements close the deployment loop for \textbf{RQ4}. Stage~1 is not only lightweight enough to co-reside on a commodity IIoT sensor node, but also fast enough to process the CSI stream in real time while preserving the SRAM and CPU budgets required by the radio and application workloads already running on the same microcontroller.

\end{document}